\documentclass[aps,prd,preprint,tightenlines,groupedaddress,nofootinbib,showpacs]{revtex4}
\usepackage{amssymb,latexsym}
\usepackage{amsmath,amsbsy}
\usepackage{epsfig,bm}
\usepackage{graphicx,comment}
\DeclareMathOperator{\tr}{tr}
\DeclareMathOperator{\Erfc}{Erfc}
\DeclareMathOperator{\Erf}{Erf}

\begin{document}
\def\a{{\alpha}}
\def\b{{\beta}}
\def\d{{\delta}}
\def\D{{\Delta}}
\def\e{{\varepsilon}}
\def\g{{\gamma}}
\def\G{{\Gamma}}
\def\k{{\kappa}}
\def\l{{\lambda}}
\def\L{{\Lambda}}
\def\m{{\mu}}
\def\n{{\nu}}
\def\o{{\omega}}
\def\O{{\Omega}}
\def\S{{\Sigma}}
\def\s{{\sigma}}
\def\th{{\theta}}

\def\ol#1{{\overline{#1}}}

\def\Dslash{D\hskip-0.65em /}
\def\Dtslash{\tilde{D} \hskip-0.65em /}

\def\CPT{{$\chi$PT}}
\def\QCPT{{Q$\chi$PT}}
\def\PQCPT{{PQ$\chi$PT}}
\def\tr{\text{tr}}
\def\str{\text{str}}
\def\diag{\text{diag}}
\def\order{{\mathcal O}}

\def\cC{{\mathcal C}}
\def\cB{{\mathcal B}}
\def\cT{{\mathcal T}}
\def\cQ{{\mathcal Q}}
\def\cL{{\mathcal L}}
\def\cO{{\mathcal O}}
\def\cA{{\mathcal A}}
\def\cQ{{\mathcal Q}}
\def\cR{{\mathcal R}}
\def\cH{{\mathcal H}}
\def\cW{{\mathcal W}}
\def\cM{{\mathcal M}}
\def\cD{{\mathcal D}}
\def\cN{{\mathcal N}}
\def\cP{{\mathcal P}}
\def\cK{{\mathcal K}}
\def\Qt{{\tilde{Q}}}
\def\Dt{{\tilde{D}}}
\def\St{{\tilde{\Sigma}}}
\def\cBt{{\tilde{\mathcal{B}}}}
\def\cDt{{\tilde{\mathcal{D}}}}
\def\cTt{{\tilde{\mathcal{T}}}}
\def\cMt{{\tilde{\mathcal{M}}}}
\def\At{{\tilde{A}}}
\def\cNt{{\tilde{\mathcal{N}}}}
\def\cOt{{\tilde{\mathcal{O}}}}
\def\cPt{{\tilde{\mathcal{P}}}}
\def\cI{{\mathcal{I}}}
\def\cJ{{\mathcal{J}}}
\def\phit{{\tilde{\phi}}}

\def\eqref#1{{(\ref{#1})}}

\preprint{UMD-40762-430}

\title{Flavor Twisted Boundary Conditions and the Nucleon Vector Current}

\author{F.-J.~Jiang}
\email[]{fjjiang@itp.unibe.ch}
\affiliation{%
Institute for Theoretical Physics,
Bern University,
Sidlerstrasse 5,
CH-3012 Bern, 
Switzerland
}

\author{B.~C.~Tiburzi}
\email[]{bctiburz@umd.edu}
\affiliation{%
Maryland Center for Fundamental Physics, 
Department of Physics, 
University of Maryland, 
College Park,  
MD 20742-4111, 
USA
}

\date{\today}


\begin{abstract}
Using flavor twisted boundary conditions, 
we study nucleon matrix elements of the vector current. 
We twist only the active quarks that couple to the current.
Finite volume corrections due to twisted boundary
conditions are determined using partially twisted, partially quenched,
heavy baryon chiral perturbation theory, which we develop 
for the graded group $SU(7|5)$. 
Asymptotically these corrections are exponentially small in the volume,
but can become pronounced for small twist angles.
Utilizing the Breit frame does not mitigate volume corrections to nucleon vector current matrix elements.
The derived expressions will allow for better controlled extractions of
the isovector magnetic moment and the electromagnetic radii from simulations at zero lattice momentum.
Our formalism, moreover, can be applied to any nucleon matrix elements. 
\end{abstract}

\maketitle

\section{Introduction}

Understanding QCD in the strongly interacting regime
remains a challenging problem in physics. 
Simulations of QCD on Euclidean spacetime lattices
are making progress towards a quantitative understanding
of the non-perturbative dynamics in QCD~\cite{DeGrand:2006aa}. 
Lattice QCD simulations usually employ periodic boundary
conditions for the quark and gluon fields. 
Consequently the available hadron momenta are 
limited to periodic momentum modes of the lattice, 
$\bm{k} = 2 \pi \bm{n} / L$, where $\bm{n}$ is 
a triplet of integers and $L$ is the lattice size in each of the
three spatial directions.   
On typical lattices, the smallest available lattice 
momentum is about $400 - 500 \, \texttt{MeV}$. 
This presents a severe limitation for the study of observables 
appearing in matrix elements at 
low momentum, and low momentum-transfer.
At present, such observables cannot be investigated directly
using periodic boundary conditions, 
and models are used to perform momentum extrapolations.

For large enough volume, 
the physics should be independent of the choice of boundary conditions.
There is freedom in choosing boundary conditions for fields; however, 
the action must be single valued so that observables are well-defined.
For a generic matter field $\phi$,
we can impose a twisted boundary condition in the 
$i$-th direction of the form, see e.g.~\cite{ZinnJustin:2002ru},
\begin{equation}  \notag
\phi(x_i + L) = U \, \phi(x_i)
,\end{equation}
where $U$ is a symmetry of the action and $U^\dagger U = 1$. 
For the quark flavors in QCD, 
the diagonal flavor rotations can be used
to implement what are called flavor-twisted
boundary conditions. 
With $U$ of the form $U = \exp ( i \theta_i)$,
the matter field $\phi$ has kinematic momentum
$\bm{k} = (2 \pi \bm{n} + \bm{\theta}) /  L$
which can be varied continuously by choosing 
different values for $\bm{\theta}$. 
The ability to produce continuous hadron momentum
has made flavor-twisted boundary conditions attractive to 
lattice QCD~\cite{Bedaque:2004kc,deDivitiis:2004kq,Sachrajda:2004mi,Bedaque:2004ax,Tiburzi:2005hg,Mehen:2005fw,Flynn:2005in,Guadagnoli:2005be,Aarts:2006wt,Tiburzi:2006px,Jiang:2006gna,Boyle:2007wg,Simula:2007fa,Boyle:2008yd,Jiang:2008te,Aoki:2008gv}.

In this work, 
we detail the finite volume modifications to nucleon form factors of the vector current. 
We use twisted boundary conditions on the active quarks in the current insertion,
and, of course, are limited to only connected contributions from the current. 
Heavy baryon chiral perturbation theory is utilized to estimate the volume
dependence of nucleon current matrix elements. 
Let us summarize our main findings. 
\begin{itemize}
\item
Finite volume modifications can be sizable especially for the magnetic 
contribution, and for small twist angles.
\item
The use of Breit frame kinematics does not dramatically reduce or 
simplify the finite volume corrections. 
The volume effect for magnetic observables in the Breit frame roughly
doubles compared to the rest frame. 
This situation is unlike the meson vector current~\cite{Jiang:2008te}.
\item
With twisted boundary conditions on only the active quarks, 
the finite volume corrections depend on an unphysical and unknown parameter,
$g_1$.
This dependence arises as an artifact of the enlarged valence flavor group, 
and a lattice determination of $g_1$ would help in accounting for volume 
corrections. 
\item
Results obtained here are qualitatively similar to those obtained 
from isospin-twisted boundary conditions for the nucleon 
isovector form factors~\cite{Tiburzi:2006px}.
In that method, valence $u$-quarks are twisted differently than the valence $d$-quarks
without introducing extra fictitious flavors.  
We show, moreover, the flavor symmetry employed by that method
can be used to eliminate the dependence on $g_1$. 
Consequently finite volume corrections can be reliably estimated for that case
in terms of known low-energy constants.
\end{itemize}

Our presentation has the following organization. 
First in Section~\ref{s:tBXPT}, 
we detail the flavor twisted boundary conditions, and incorporate them into heavy baryon chiral perturbation theory. 
We show how the graded group $SU(7|5)$ accommodates twisting of the active valence quarks in the baryon sector. 
In Section~\ref{s:mass}, 
we compute finite volume corrections to the nucleon mass,
and derive the induced mass splittings due to flavor twisted boundary conditions. 
Numerically these splittings are estimated to be at the percent level or less on current lattices.
Finite volume corrections to the vector form factors of the proton and neutron 
are determined in Section~\ref{s:vec} (complete expressions are given in Appendix~\ref{A:result}). 
These results apply to the connected contributions allowing access to isovector quantities, but not isoscalar. 
We show that terms arising from broken cubic invariance can lead to non-negligible volume effects in the region of small twist angles.
Results for rest frame and Breit frame kinematics are compared. 
The Breit frame does not offer any substantial advantages with respect to volume effects.
Complete results for isovector current matrix elements at finite volume
using the method of isospin twisted boundary conditions are displayed in Appendix~\ref{A:oldresult}. 
For rest frame kinematics, these results are shown to be independent of the unphysical 
parameter $g_1$. 
Appendix~\ref{s:FVS} collects functions and identities useful for the evaluation of finite volume effects. 
Finally a brief summary concludes our work.

\section{Flavor twisted boundary conditions and baryon chiral perturbation theory} \label{s:tBXPT}

To address the consequences of twisted boundary conditions in lattice calculations of baryon properties, 
we describe the underlying effective theory in the baryon sector. 
First we detail the partially twisted boundary conditions employed. 
Next we include these effects in chiral perturbation theory,
and then heavy baryon chiral perturbation theory.

The quark part of the partially quenched QCD Lagrangian is given by
\begin{equation}
\mathcal{L} = \sum_{j,k=1}^{12} \ol{ \hat{Q} } \,  {}^{j} 
\left(
\Dslash +  m_Q \right)_j {}^{k}  \,
\hat{Q}_k
.\label{eq:pqqcdlag}
\end{equation}
The twelve quark fields transform in the fundamental representation of the graded 
group $SU(7|5)$, 
and appear in the vector 
$\hat{Q}^{\text{T}} = (\hat{u}_0, \hat{u}_1,\hat{u}_2, \hat{d}_1, \hat{d}_2, \hat{j}, \hat{l}, \hat{\tilde{u}}_0, \hat{\tilde{u}}_1, \hat{\tilde{u}}_2, \hat{\tilde{d}}_1, \hat{\tilde{d}}_2)$.
In addition to the valence
$\hat{u}_0$, $\hat{u}_1$, $\hat{u}_2$, $\hat{d}_1$, 
and 
$\hat{d}_2$ 
quarks, 
we have added their ghost quark counterparts
$\hat{\tilde{u}}_0$, $\hat{\tilde{u}}_1$, $\hat{\tilde{u}}_2$, $\hat{\tilde{d}}_1$ 
and 
$\hat{\tilde{d}}_2$, 
which cancel the closed valence loops, 
and two sea quarks 
$\hat{j}$ 
and 
$\hat{l}$. 
In the isospin limit, 
the quark mass matrix of 
$SU(7|5)$ 
reads
$m_Q = \diag(m_u \bm{1}_{5 \times 5},  m_j \bm{1}_{2 \times 2},  m_u \bm{1}_{5 \times 5})$
in block diagonal form, where the blocks correspond to valence, sea and ghost sectors.
QCD quantities can be recovered in the limit $m_j \rightarrow m_u$.
The additional up and down quarks are fictitious flavors differing only by their boundary conditions. 
There is one more up-type quark than down-type quark because we focus on a theory
that will yield proton matrix elements. Neutron matrix elements can always be derived
trivially by interchanging up and down charges in the final result.%
\footnote{
To consider both proton and neutron properties in the same 
theory, we would need to enlarge the flavor group further to $SU(8|6)$.
}

The hats denote fields satisfying twisted boundary conditions.
We require that the quark fields satisfy boundary conditions of the form
\begin{equation}
\hat{Q}(x + L \bm{e}_r) = \exp \left( i \theta^a_r \, \ol T {}^a \right) \hat{Q}(x)  
,\end{equation}
where 
$\bm{e}_r$ 
is a unit vector in the 
$r^{\text{th}}$ 
spatial direction, 
$L$ 
is the spatial size of the lattice,
and the block diagonal form of the supermatrices 
$\ol T {}^a$ 
is
\begin{equation} \label{eq:qtwist}
\ol T {}^a 
= 
\diag 
\left( T^a, 0, T^a \right)
.\end{equation} 
Here we choose
$T^a$ 
to be generators of the
$U(5)$ 
Cartan subalgebra. 
Notice that by Eq.~\eqref{eq:qtwist}, 
the sea quarks remain periodic at the boundary. 
This reflects a partially twisted scenario. 
Twist angles can be changed without necessitating the generation of new gauge configurations, 
because the fermionic determinant, which arises solely from the sea sector, is not affected by the twisting.

Redefining the quark fields as 
$Q^{\text{T}} = (u_0, u_1, u_2, d_1, d_2, j, l, \tilde{u}_0, \tilde{u}_1, \tilde{u}_2, \tilde{d}_1, \tilde{d}_2)$,
with
$Q(x) = V^\dagger(x) \hat{Q}(x)$, 
where 
$V(x) = \exp ( i \bm{\th}^a \cdot  \bm{x} \, \ol T {}^a / L )$, 
we can write the partially quenched QCD Lagrangian as
\begin{equation}
\cL 
= 
\sum_{j,k=1}^{12} \ol{Q} \, {}^{j} 
\left(
\hat{\Dslash} + m_Q 
\right)_j {}^{k} \, 
Q_k
,\end{equation}
where all 
$Q$ 
fields satisfy periodic boundary conditions, and the effect of twisting has the form of a uniform gauge field:  
$\hat{D}_\mu = D_\mu + i B_\mu$, 
where 
$B_\mu = (\bm{\th}^a \, \ol T {}^a / L , 0)$. 
It will be easier to treat the twisting in the flavor basis of the valence and ghost sectors rather than in the generator basis, 
thus we write
$\bm{\th}^a \, T {}^a = \diag (\bm{0}, \bm{\th}^u, \bm{\th'} {}^{u}, \bm{\th}^d, \bm{\th'} {}^{d}  )$, 
and similarly for $B_\mu$, which appears as
$B_\mu = \diag (B_\mu^{\text{val}}, 0,  B_\mu^{\text{val}} )$
in block diagonal form, with
$B_\mu^{\text{val}} = \diag ( 0, B^u_\mu, B^{'u}_\mu, B^d_\mu, B^{'d}_\mu )$. 
Momentum transfer will be generated using flavor changing currents
from $u_1$ to $u_2$, or from $d_1$ to $d_2$. 
Notice we keep the $u_0$ quark periodic; 
it will play the role of spectator.

The low-energy effective theory of QCD is chiral perturbation theory, 
which describes the dynamics of pseudoscalar mesons emerging from spontaneous
chiral symmetry breaking. 
The mesons of partially quenched chiral perturbation theory%
~\cite{Bernard:1994sv,Sharpe:1997by,Golterman:1998st,Sharpe:2000bc,Sharpe:2001fh}
are described by the coset field 
$\hat{\Sigma}$, 
which satisfies twisted boundary conditions.
This field can be traded in for  
$\S$, 
defined by
$\S (x) = V^\dagger(x) \hat{\S} (x) V(x)$,
which is periodic at the boundary~\cite{Sachrajda:2004mi}. 
In terms of this field, 
the Lagrangian of partially quenched chiral perturbation theory appears as
\begin{equation} \label{eqn:Lchi}
\cL = 
\frac{f^2}{8} \str \left( \hat{D}_\mu \S \hat{D}_\mu \S^\dagger \right)
- 
\l \, \str \left( m_Q^\dagger \S + \S^\dagger m_Q \right)
+
\mu_0^2 \Phi_0^2
.\end{equation}
The action of the covariant derivative 
$\hat{D}^\mu$ 
is specified by
$\hat{D}_\mu \S = D_\mu \S + i [ B_\mu, \S ]$.
The parameter $f$ is the chiral limit value of the pion 
decay constant, and in our normalization, $f = 0.13 \, \texttt{GeV}$. 
The above Lagrangian contains only the lowest-order terms
in an expansion in quark mass $m_Q$, and meson momentum $k^2$. 
The periodic meson fields contained in the twelve-by-twelve matrix $\phi$  
are realized nonlinearly,
$\S = \exp ( 2 i \phi / f)$. 
The matrix $\phi$ has the form
\begin{equation}
\phi
=
\begin{pmatrix}
M_{vv} & M_{vs} & \chi^\dagger_{gv} \\
M_{sv} & M_{ss} & \chi^\dagger_{gs} \\
\chi_{gv} &  \chi_{gs} & M_{gg} 
\end{pmatrix}
.\end{equation}
The mesons of $M_{vv}$ ($M_{gg}$)  are bosonic and 
are formed from a valence (ghost) quark-antiquark pair. 
These matrices have the form
\begin{equation}
M_{vv}
=
\begin{pmatrix}
\eta^u_{00}  & \eta^u_{01}  & \eta^u_{02} & \pi^+_{01}   & \pi^+_{02} \\
\eta^u_{10}  & \eta^u_{11}  & \eta^u_{12} & \pi^+_{11} & \pi^+_{12} \\
\eta^u_{20}  & \eta^u_{21}  & \eta^u_{22} & \pi^+_{21} & \pi^+_{22} \\
\pi^-_{10}     & \pi^-_{11}     & \pi^-_{12} & \eta^d_{11} & \eta^d_{12} \\
\pi^-_{20}  & \pi^-_{21} & \pi^-_{22} & \eta^d_{21} & \eta^d_{22}
\end{pmatrix}
, \, \,
\text{and}
\quad
M_{gg}
=
\begin{pmatrix}
\tilde{\eta}^u_{00}  & \tilde{\eta}^u_{01}  & \tilde{\eta}^u_{02} & \tilde{\pi}^+_{01}   & \tilde{\pi}^+_{02} \\
\tilde{\eta}^u_{10}  & \tilde{\eta}^u_{11}  & \tilde{\eta}^u_{12} & \tilde{\pi}^+_{11}  & \tilde{\pi}^+_{12} \\
\tilde{\eta}^u_{20}  & \tilde{\eta}^u_{21}  & \tilde{\eta}^u_{22} & \tilde{\pi}^+_{21} & \tilde{\pi}^+_{22} \\
\tilde{\pi}^-_{10}     & \tilde{\pi}^-_{11} & \tilde{\pi}^-_{12} & \tilde{\eta}^d_{11} & \tilde{\eta}^d_{12} \\
\tilde{\pi}^-_{20}  & \tilde{\pi}^-_{21} & \tilde{\pi}^-_{22} & \tilde{\eta}^d_{21} & \tilde{\eta}^d_{22}
\end{pmatrix}
\notag
.\end{equation}
The 
$\eta^q_{ij}$ ($\tilde{\eta}^q_{ij}$)  
mesons have quark content 
$\eta^q_{ij} \sim  q_i  \ol q_j$
($\tilde{\eta}^q_{ij} \sim  \tilde{q}_i  \overline{\tilde{q}}_j$), 
while the 
$\pi^+_{ij}$ 
($\tilde{\pi}^+_{ij}$)
mesons have quark content
$\pi^+_{ij} \sim u_i \ol d_j$ 
($\tilde{\pi}^+_{ij} \sim \tilde{u}_i \overline{\tilde{d}}_j$ ).
The valence-sea (sea-sea) mesons are bosonic and contained in 
$M_{vs}$ 
($M_{ss}$)
as
\begin{equation}
M_{sv}
=
\begin{pmatrix}
\phi_{j u_0} & \phi_{j u_1}  & \phi_{j u_2} & \phi_{j d_1} & \phi_{j d_2} \\
\phi_{l u_0} & \phi_{l u_1}  & \phi_{l u_2} & \phi_{l d_1} & \phi_{l d_2}
\end{pmatrix}
, \,\,
\text{and} 
\quad
M_{ss} 
=
\begin{pmatrix}
\eta_j & \pi_{jl} \\
\pi_{lj} & \eta_l
\end{pmatrix}
\notag
.\end{equation}
Mesons contained in $\chi_{gv}$ ($\chi_{gs}$)
are built from ghost quark, valence antiquark (sea antiquark) pairs
and are thus fermionic. These states appear as
\begin{equation}
\chi_{gv}
=
\begin{pmatrix}
\phi_{\tilde{u}_0 u_0} &
\phi_{\tilde{u}_0 u_1} & 
\phi_{\tilde{u}_0 u_2} & 
\phi_{\tilde{u}_0 d_1} & 
\phi_{\tilde{u}_0 d_2} \\ 
\phi_{\tilde{u}_1 u_0} &
\phi_{\tilde{u}_1 u_1} & 
\phi_{\tilde{u}_1 u_2} & 
\phi_{\tilde{u}_1 d_1} & 
\phi_{\tilde{u}_1 d_2} \\ 
\phi_{\tilde{u}_2 u_0} &
\phi_{\tilde{u}_2 u_1} & 
\phi_{\tilde{u}_2 u_2} &
\phi_{\tilde{u}_2 d_1} & 
\phi_{\tilde{u}_2 d_2} \\
\phi_{\tilde{d}_1 u_0} &
\phi_{\tilde{d}_1 u_1} & 
\phi_{\tilde{d}_1 u_2} & 
\phi_{\tilde{d}_1 d_1} & 
\phi_{\tilde{d}_1 d_2} \\
\phi_{\tilde{d}_2 u_0} &
\phi_{\tilde{d}_2 u_1} & 
\phi_{\tilde{d}_2 u_2} & 
\phi_{\tilde{d}_2 d_1} & 
\phi_{\tilde{d}_1 d_2} 
\end{pmatrix}
, \, \, 
\text{and}
\quad
\chi_{gs}
=
\begin{pmatrix}
\phi_{\tilde{u}_0 j} &
\phi_{\tilde{u}_0 l} \\
\phi_{\tilde{u}_1 j} & 
\phi_{\tilde{u}_1 l} \\ 
\phi_{\tilde{u}_2 j} & 
\phi_{\tilde{u}_2 l} \\ 
\phi_{\tilde{d}_1 j} & 
\phi_{\tilde{d}_1 l} \\ 
\phi_{\tilde{d}_2 j} & 
\phi_{\tilde{d}_2 l} 
\end{pmatrix}
\notag
.\end{equation}

Expanding the Lagrangian in 
Eq.~\eqref{eqn:Lchi} to lowest order, 
one finds that mesons with quark content 
$Q\bar{Q'}$
have mass
\begin{equation}\label{eqn:mqq}
m_{QQ'}^2
=
\frac{4\l}{f^2}  (m_Q+m_{Q'})
.\end{equation}
Thus in infinite volume 
all mesons fall into one of three groups of mass degenerate states:
valence-valence pions
$m_\pi^2 = 8 \l m_u / f^2$, 
valence-sea mesons
$m_{ju}^2 = 4 \l (m_u + m_j) / f^2$, 
and 
sea-sea pions
$m_{jj}^2 = 8 \l m_j / f^2$.
In partially quenched simulations, one measures 
the valence-valence and sea-sea pion masses. 
The valence-sea mass is given by the average of the other two, up to possible discretization errors
that arise in hybrid actions.

The flavor singlet field, 
$\Phi_0 =  \frac{1}{\sqrt{2}} \str  \, \phi $, 
additionally acquires a mass 
$\mu_0$
which arises as a consequence of the $U(1)_A$ anomaly. 
Taking this mass to be large, 
the flavor singlet field is then integrated out of partially quenched chiral perturbation theory, 
however,
the propagators of the flavor-neutral fields deviate from simple pole forms%
~\cite{Sharpe:2000bc,Sharpe:2001fh}. 
There are two useful simplifications to note: 
twisted boundary conditions have no effect on the flavor neutral sector,
and 
all valence-valence flavor neutral states are degenerate with mass $m_\pi$. 
For $a$,$b = u_0$, $u_1$, $u_2$,  $d_1$,  or $d_2$, 
the leading-order $\eta_a \eta_b$ propagator is thus given by
\begin{equation}
{\cal G}_{\eta_a \eta_b} 
=
\delta_{ab} \frac{1}{k^2 + m_\pi^2}
- 
\frac{1}{2} 
\frac{k^2 + m^2_{jj} }{ ( k^2 + m^2_\pi)^2 }
.\end{equation}
The flavor neutral propagator can be conveniently rewritten as
\begin{equation}
{\cal G}_{\eta_a \eta_b} 
=
\frac{  \d_{ab} } {k^2 + m_\pi^2} 
+
{\cal H}_{ab} \left( \frac{1}{k^2 + m_\pi^2} \right)
,\end{equation}
where
\begin{eqnarray}
{\cal H}_{ab}
\left(A\right) 
&=& 
-\frac{1}{2}
\left[ 
1 + (m_\pi^2 - m_{jj}^2 ) \frac{\partial}{\partial m_\pi^2} 
\right] A \, .
\label{eq:neutral}
\end{eqnarray}

To include baryons into partially quenched chiral perturbation theory, 
one uses rank three flavor tensors \cite{Labrenz:1996jy,Savage:2001dy,Chen:2001yi,Beane:2002vq}.
In $SU(7|5)$, 
the spin-$\frac{1}{2}$ baryons are described by the $\bm{572}$-dimensional supermultiplet $\hat{\cB}^{ijk}$, 
while the spin-$\frac{3}{2}$ baryons are described by the $\bm{300}$-dimensional supermultiplet $\hat{\cT}_\mu^{ijk}$%
~\cite{Tiburzi:2005is}. 
The baryon flavor tensors are twisted at the boundary of the lattice. 
In the $r^{\text{th}}$ spatial direction, both tensors satisfy boundary conditions of the form~\cite{Tiburzi:2005hg}
\begin{eqnarray}
\hat{\cB}_{ijk} (x + \hat{\bm{e}_r} L ) 
&=&
\left( e^{i \theta^a_r \, \ol T {}^a }\right)_{ii}  
\left( e^{i \theta^a_r \, \ol T {}^a }\right)_{jj}  
\left( e^{i \theta^a_r \, \ol T {}^a }\right)_{kk} 
\hat{\cB}_{ijk}(x)
.\end{eqnarray}
Thus we define new tensors $\cB^{ijk}$ and $\cT_\mu^{ijk}$ both having the form
\begin{equation}
\cB_{ijk}(x) = V^\dagger_{ii}(x) V^\dagger_{jj}(x) V^\dagger_{kk}(x) \hat{\cB}_{ijk}(x)
.\end{equation}
These baryon fields satisfy periodic boundary conditions, 
and their free Lagrangian has the form
\begin{eqnarray}
  \cL
  &=&
   -i\left(\ol\cB v_\mu \hat{\cD}_\mu  \cB \right)
  -2\a_M^{(PQ)} \left(\ol \cB \cB \cM_+ \right)
  -2\b_M^{(PQ)} \left(\ol\cB \cM_+\cB \right)
  -2\sigma_M^{(PQ)} \left(\ol\cB \cB \right) \str\left( \cM_+\right)
                              \nonumber \\
  &&
  -i \left(\ol\cT_\nu v_\mu \hat{\cD}_\mu  \cT_\nu\right)
  + \D\left(\ol\cT_\nu \cT_\nu \right)
  +2\g_M^{(PQ)}\left(\ol\cT_\nu \cM_+ \cT_\nu \right)
  +2\ol\sigma_M^{(PQ)}\left(\ol\cT_\nu\cT_\nu\right)\str\left( \cM_+ \right) 
,\notag \\ 
 \label{eq:LB}
 \end{eqnarray}
where 
$v_\mu = ( 0, 0, 0, i)$ 
is the Euclidean four-velocity in the rest frame. 
The mass operator 
$\cM_+$ 
is defined by 
$\cM_+ = \frac{1}{2}\left(\xi^\dagger m_Q \xi^\dagger + \xi m_Q \xi \right)$, 
with 
$\xi = \sqrt{\S}$, 
and the covariant derivative acts on 
$\cB$ 
and 
$\cT_\mu$ fields in the same manner, 
namely
\begin{equation}
[\hat{\cD}_\mu \cB (x)]^{ijk} 
= 
[\cD_\mu \cB]^{ijk} (x) + i (B_\mu^i + B_\mu^j + B_\mu^k) \cB^{ijk}(x)
,\end{equation}
with
\begin{equation}
[\cD_\mu \cB]^{ijk}
= 
\partial_\mu \cB^{ijk}
+ 
(\hat{V}_\mu)^{il} 
\cB^{ljk}
+
(-)^{\eta_i (\eta_j + \eta_l) }
(\hat{V}_\mu)^{jl} 
\cB^{ilk}
+
(-)^{(\eta_i + \eta_j)(\eta_k + \eta_l)} 
(\hat{V}_\mu)^{kl} 
\cB^{ijl}
,\end{equation}
where the vector field of mesons $\hat{V}_\mu$ is given by
$\hat{V}_\mu = \frac{1}{2} \left( \xi \hat{D}_\mu \xi^\dagger + \xi^\dagger \hat{D}_\mu \xi \right)$. 
This free Lagrangian contains a number of low-energy constants
but has precisely the same form as in the $SU(4|2)$ partially quenched theory.
Restricting the baryon multiplets to the sea sector, 
so that all flavor indices are either $6$ or $7$, 
we have nucleons and deltas made only of sea quarks. 
Hence the matching conditions are precisely the same 
as those used to match $SU(4|2)$ onto $SU(2)$ by 
restricting the former to the sea sector. 
The relations between the low-energy constants appearing in Eq.~\eqref{eq:LB} 
and those of 
$SU(2)$ 
chiral perturbation theory will not be needed here, but are given in~\cite{Beane:2002vq}.

The leading order partially quenched interaction Lagrangian 
between the baryons and mesons appears as
\begin{equation} \label{eq:Lint}
\cL 
= 
2 \a \left( \ol \cB S_\mu \cB \hat{A}_\mu \right)
+ 
2 \b \left( \ol \cB S_\mu \hat{A}_\mu \cB \right)
- 
2 \cH \left( \ol \cT_\nu S_\mu \hat{A}_\mu \cT_\nu \right) 
+  
\sqrt{\frac{3}{2}} \cC 
\left[ 
\left( \ol \cT_\nu \hat{A}_\nu \cB \right)
+ 
\left( \ol \cB \hat{A}_\nu \cT_\nu \right)
\right]
,\end{equation}
where the effects of partial twisting show up in the  axial-vector field of mesons 
$\hat{A}_\mu = \frac{i}{2} \left( \xi \hat{D}_\mu \xi^\dagger - \xi^\dagger \hat{D}_\mu \xi \right)$.
The interaction Lagrangian has the same form as the $SU(4|2)$ theory of baryons, 
hence the matching conditions to $SU(2)$ are identical. 
The familiar low-energy constants of $SU(2)$ are identified as follows~\cite{Beane:2002vq}: 
$g_A = \frac{2}{3} \a - \frac{1}{3} \b$, $g_{\D N} = - \cC$, and $g_{\D \D} = \cH$. 
Notice there is an extra free parameter in the partially quenched  interaction Lagrangian compared to that of $SU(2)$ chiral peturbation theory.
We shall write our expressions in terms of $g_A$ and the combination $g_1 = \frac{1}{3} \a + \frac{4}{3} \b$. 
Dependence on $g_1$ must drop out in the QCD limit, 
which, 
for the case at hand, 
requires both $m_j \to m_u$ and $L \to \infty$.

\section{Nucleon Mass} \label{s:mass}

To begin, 
we determine the nucleon mass in the presence of partially twisted boundary conditions. 
In the isospin limit of $SU(4|2)$, 
the proton and neutron are degenerate~\cite{Beane:2002vq,Tiburzi:2005na}. 
Flavor twisted boundary conditions, 
however, 
break the valence flavor symmetry,
hence the nucleons are no longer degenerate. 
The nucleon mass splittings arise from finite volume effects induced by the boundary conditions. 
Effects of this type can be treated using chiral perturbation theory at finite volume.
We work in the $p$-regime throughout, where $m_\pi L \gg 1$ so that zero modes of the pion 
field do not become strongly coupled~\cite{Gasser:1987ah,Leutwyler:1987ak,Gasser:1987zq}.
We estimate the size of the mass splittings on current-sized lattices, 
and show that their effect can be neglected for the determination of nucleon observables
using twisted boundary conditions.

In the infinite volume limit with $B_\mu$ held fixed, 
the nucleon mass is unaffected by the boundary conditions. 
This follows from a generalization of the argument presented for mesons in~\cite{Sachrajda:2004mi}.
One merely realizes that the nucleon propagators are not boosted in heavy baryon chiral perturbation theory, because 
$v_\mu  B_\mu = 0$.
The remaining momenta in a given diagram are mesonic, and are boosted according to flavor. 
Now since there are no flavor changing interactions, the sum of boosts at each vertex is zero.
This, along with $v_\mu B_\mu = 0$, assures us we can always shift loop momenta to cast any diagram into a form where 
the only $B$-dependence is that from external momenta. 
These contributions should be thought of as kinematical rather than effects which arise in the loops from chiral dynamics.  
The mass does not receive dynamical corrections from the boundary conditions, 
but the energy depends on the external momentum and has a kinematic dependence on the boundary
conditions, e.g.~for a nucleon with lattice momentum $\bm{k}$ 
\begin{equation}
E_N = M_N + \frac{(\bm{k} + \bm{B})^2}{2 M_N} + \ldots
,\end{equation}
where $\bm{B} = \bm{\theta} / L$, and $\ldots$ denotes terms that are higher order in $1/M_N$.
Here only one valence quark in the nucleon has been twisted, and by an angle $\bm{\theta}$. 
When we try to apply the same reasoning at finite volume, no shifts of the internal momenta 
are possible because the loop momenta are discrete, while the twisting parameters are continuous.
Thus there is a dynamical dependence on the twisting parameters arising from loops,
and in finite volume, $M_N$ will depend upon $\bm{B}$.

\begin{figure}
\begin{center}
\epsfig{file=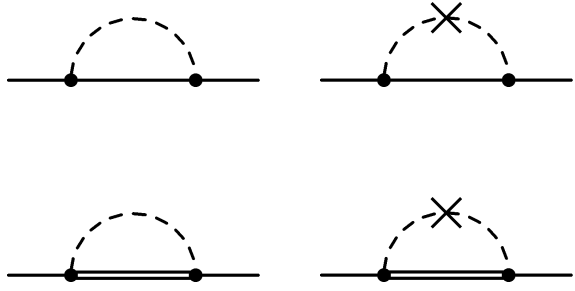,width=3in}
\caption{Diagrams contributing to the nucleon mass and wavefunction renormalization in partially quenched chiral perturbation theory.
A thin (thick) line denotes a spin-$1/2$ (spin-$3/2$) baryon, while a dashed line denotes
a meson. Partially quenched hairpins are depicted by a crossed dashed line. 
}
\label{f:Nmass}
\end{center}
\end{figure}

The mass of the nucleon in the chiral expansion can be written in the form
\begin{equation}
M_{N} = M_0(\mu) - M_N^{(1)}(\mu) - M_{N}^{(3/2)}(\mu) + \ldots
,\end{equation}
where $\mu$ is the renormalization scale, and $M_N^{(n)}$ denotes the contribution to
the nucleon mass of order $m_q^{n}$. The linear quark mass dependence arises
from the local operators in Eq.~\eqref{eq:LB} at tree level, while the leading non-analytic 
contribution $\cO(m_q^{3/2})$ arises from the one-loop diagrams shown in Figure~\ref{f:Nmass}.
The local interactions do not contribute to finite volume effects, only the meson loops that are shown in the figure. 
For periodic boundary conditions, the finite volume effects on the nucleon mass have been determined 
in~\cite{Beane:2004tw}. 
To express the finite volume corrections to the nucleon mass with flavor twisted boundary conditions, 
we require the mode sum
\begin{equation} \label{eq:Kfunc}
\cK (m,\bm{B}, \D)  
= 
\int_0^\infty d \l 
\left[ 
\frac{1}{L^3}
\sum_{\bm{n}} 
\frac{(\bm{k} + \bm{B})^2}{[(\bm{k} + \bm{B})^2 + \b_\D^2]^{3/2}}
-
\int \frac{d\bm{k}}{(2 \pi)^3} 
\frac{\bm{k}^2}{[\bm{k}^2 + \b_\D^2]^{3/2}}
\right]
,\end{equation}
with $\b_\D^2 = \l^2 + 2 \D \l + m^2$, and $\bm{k} = 2 \pi \bm{n} / L$ where $\bm{n}$ is a triplet of integers.  
Evaluation of this function, as well as other finite volume sums, is discussed in Appendix~\ref{A:result}.

Consider purely valence nucleon states with exactly one twisted quark. 
These will be the only nucleons relevant in the computation of matrix elements with twisted boundary conditions. 
It is easiest to classify these states according to their representations under the valence subgroup of the 
two degenerate untwisted quarks. In our formulation, we must set the twist angles for these quarks to zero by hand. 
There is both a singlet, $\bm{1}$, and triplet, $\bm{3}$, representation for singly twisted nucleons under the untwisted valence
$SU(2)$.  
For the mass of  a nucleon with one twisted valence quark in the $\bm{3}$ representation, 
we find the finite volume shift 
$\d M_{N_{\bm{3}} } (\bm{B})$ 
is given by
\begin{eqnarray}
\d M_{N_{\bm{3}} } (\bm{B})
&=&
-
\frac{1}{2 f^2}
\Bigg\{
g_{\pi N_{\bm{3}} N_{\bm{3}}}^2 
 \cK(m_{\pi}, \bm{0}, 0)
+ 
g_{\pi N_{\bm{3}} N_{\bm{3}}}^{\prime \,  2} 
\cK(m_\pi, \bm{B}, 0) 
+
g_{ju N_{\bm{3}} N_{\bm{3}}}^{2}  
\cK(m_{ju}, \bm{0}, 0) 
\notag \\
&& \phantom{spacer.}
+  
g_{ju N_{\bm{3}} N_{\bm{3}}}^{\prime \,  2}  
\cK(m_{ju}, \bm{B}, 0)
+  
(g_A + g_1)^2 \cH_{uu} \Big(\cK ( m_{\pi}, \bm{0}, 0) \Big)
\notag \\
&& \phantom{}
+ \frac{1}{9} g_{\D N}^2 
\Bigg[
\cK (m_{\pi}, \bm{0}, \D) 
+ 5 \cK ( m_\pi, \bm{B}, \D) 
+ 2 \cK ( m_{ju}, \bm{0}, \D) + 4 \cK (m_{ju}, \bm{B}, \D)
\Bigg]
\Bigg\} .
\notag \\
\label{eq:masshift3}
\end{eqnarray}
The effective axial couplings are defined by
\begin{eqnarray}
g_{\pi N_{\bm{3}} N_{\bm{3}}}^2 
&=& 
\frac{1}{3} (g_A^2 + 2 g_A g_1 + g_1^2 / 4),
\notag \\
g_{\pi N_{\bm{3}} N_{\bm{3}}}^{\prime \, 2} 
&=& 
\frac{1}{3} (g_A^2 -   g_A g_1 - 5 g_1^2 / 4),
\notag \\
g_{ju N_{\bm{3}} N_{\bm{3}}}^{2} 
&=&
\frac{1}{3} ( 4 g_A^2 + 2 g_A g_1 + g_1^2),
\notag \\
g_{ju N_{\bm{3}} N_{\bm{3}}}^{\prime \, 2} 
&=&
 \frac{1}{2} g_1^2.
\label{eq:N3axial}
\end{eqnarray}
On the other hand, for a nucleon in the $\bm{1}$ representation,
we find the finite volume shift $\d M_{N_{\bm{1}} } (\bm{B})$ given by
\begin{eqnarray}
\d M_{N_{\bm{1}} } (\bm{B})
&=&
-
\frac{1}{2 f^2}
\Bigg\{
g_{\pi N_{\bm{1}} N_{\bm{1}}}^2 
\cK(m_{\pi}, \bm{0}, 0)  
+ 
g_{\pi N_{\bm{1}} N_{\bm{1}}}^{\prime \, 2} 
\cK(m_\pi, \bm{B}, 0) 
+
g_{ju N_{\bm{1}} N_{\bm{1}}}^{2} 
\cK(m_{ju}, \bm{0}, 0) 
\notag \\
&& \phantom{spacer.}
+  
g_{ju N_{\bm{1}} N_{\bm{1}}}^{\prime \, 2} 
\cK(m_{ju}, \bm{B}, 0)
+ 
(g_A + g_1)^2 \cH_{uu} \Big(\cK ( m_{\pi}, \bm{0}, 0) \Big)
\notag \\ 
&& \phantom{sp}
+ \frac{1}{3} g_{\D N}^2 
\Bigg[
\cK (m_{\pi}, \bm{0}, \D) 
+ \cK ( m_\pi, \bm{B}, \D) 
+ 2 \cK ( m_{ju}, \bm{0}, \D) 
\Bigg]
\Bigg\} \label{eq:masshift1}
.\end{eqnarray}
The effective axial couplings for the singlet nucleon mass are defined by
\begin{eqnarray}
g_{\pi N_{\bm{1}} N_{\bm{1}}}^2 
&=& 
\frac{1}{9} (g_A^2 - 4  g_A g_1 -11  g_1^2 / 4),
\notag \\
g_{\pi N_{\bm{1}} N_{\bm{1}}}^{\prime \, 2} 
&=& 
\frac{1}{9} (5 g_A^2 + 7  g_A g_1 -  g_1^2 / 4) ,
\notag \\
g_{ju N_{\bm{1}} N_{\bm{1}}}^{2} 
&=&
\frac{1}{9} ( 4 g_A^2 + 2 g_A g_1 + 7 g_1^2) ,
\notag \\
g_{ju N_{\bm{1}} N_{\bm{1}}}^{\prime \, 2} 
&=&
\frac{1}{9}(8 g_A^2 + 4 g_A g_1 +  g_1^2/ 2) 
.\label{eq:N1axial}
\end{eqnarray}
In the limit $\bm{B} = \bm{0}$, 
there is no difference between the representations,
and we recover accordingly the finite volume shift of the 
partially quenched nucleon mass~\cite{Beane:2004tw}.
In the rest of this section, we work for simplicity
at the unitary mass point $m_j = m_u$, so that
$m_{ju}^2 = m_\pi^2$.

We consider nucleon splittings for two cases that are of interest in 
current matrix elements:
rest frame kinematics, and Breit frame kinematics. 
In the rest frame kinematics, the initial nucleon 
is at rest, and hence completely untwisted. 
The final nucleon has been given a boost by twisting one of the quarks by 
$\bm{\theta}$. 
In this case, there are three mass splittings among the various nucleons:
that between the $\bm{3}$ and untwisted nucleon, that between the $\bm{1}$ 
and untwisted nucleon, and that between the $\bm{3}$ and $\bm{1}$ nucleons. 
The relative change in these splittings is given by
\begin{eqnarray}
\label{eq:3split}
\D M_{\bm{3}} 
&\equiv&
\frac{M_{N_{ \bm{3} }} (\bm{B}) - M_N}{M_N} 
\\
&=&
- \frac{1}{2 f^2 M_N}
\Bigg\{
\frac{1}{3} ( g_A^2 - g_A g_1 + g_1^2 / 4 ) 
\Big[ \cK( m_\pi, \bm{B}, 0) - \cK( m_\pi , \bm{0}, 0 ) \Big]
\notag \\
&& \phantom{space}
+ g_{\D N}^2 
\Big[ \cK (m_\pi, \bm{B}, \D ) - \cK ( m_\pi, \bm{0}, \D ) \Big]
\Bigg\},
\end{eqnarray}
\begin{eqnarray}
\label{eq:1split}
\D M_{\bm{1}} 
&\equiv&
\frac{M_{N_{ \bm{1} }} (\bm{B}) - M_N}{M_N} 
\\
&=&
- \frac{1}{2 f^2 M_N}
\Bigg\{
\frac{1}{9} (13  g_A^2 +11 g_A g_1 + g_1^2 / 4 ) 
\Big[ \cK( m_\pi, \bm{B}, 0) - \cK( m_\pi , \bm{0}, 0 ) \Big]
\notag \\
&& \phantom{space}
+ \frac{1}{3} g_{\D N}^2 
\Big[ \cK (m_\pi, \bm{B}, \D ) - \cK ( m_\pi, \bm{0}, \D ) \Big]
\Bigg\},
\end{eqnarray}
and
\begin{eqnarray}
\label{eq:31split}
\D M_{\bm{3} -\bm{1}} 
&\equiv&
\frac{M_{N_{ \bm{3} }} (\bm{B}) - M_{N_{\bm{1}}}(\bm{B}) }{M_N} 
\\
&=&
- \frac{1}{2 f^2 M_N}
\Bigg\{
-
\frac{1}{9} (10  g_A^2 +14 g_A g_1 - g_1^2 / 2 ) 
\Big[ \cK( m_\pi, \bm{B}, 0) - \cK( m_\pi , \bm{0}, 0 ) \Big]
\notag \\
&& \phantom{space}
+ \frac{2}{3} g_{\D N}^2 
\Big[ \cK (m_\pi, \bm{B}, \D ) - \cK ( m_\pi, \bm{0}, \D ) \Big]
\Bigg\},
\end{eqnarray}
respectively.

On the other hand, for the Breit frame kinematics
the initial state nucleon has one quark twisted by $\bm{\theta}$, 
while the final state nucleon has one quark twisted by $- \bm{\theta}$. 
The finite volume modification given by the function $\cK ( m, \bm{B}, \D) $ in Eq.~\eqref{eq:Kfunc}
is even with respect to $\bm{B}$. 
Thus initial and final state nucleons in the same representation of the untwisted $SU(2)$ are degenerate. 
The only non-vanishing splitting is between the different representations, 
but on account of evenness in $\bm{B}$, this splitting is identical to  
$\D M_{\bm{3} - \bm{1}}$ given above in Eq.~\eqref{eq:31split}.

\begin{figure}
\begin{center}
\epsfig{file=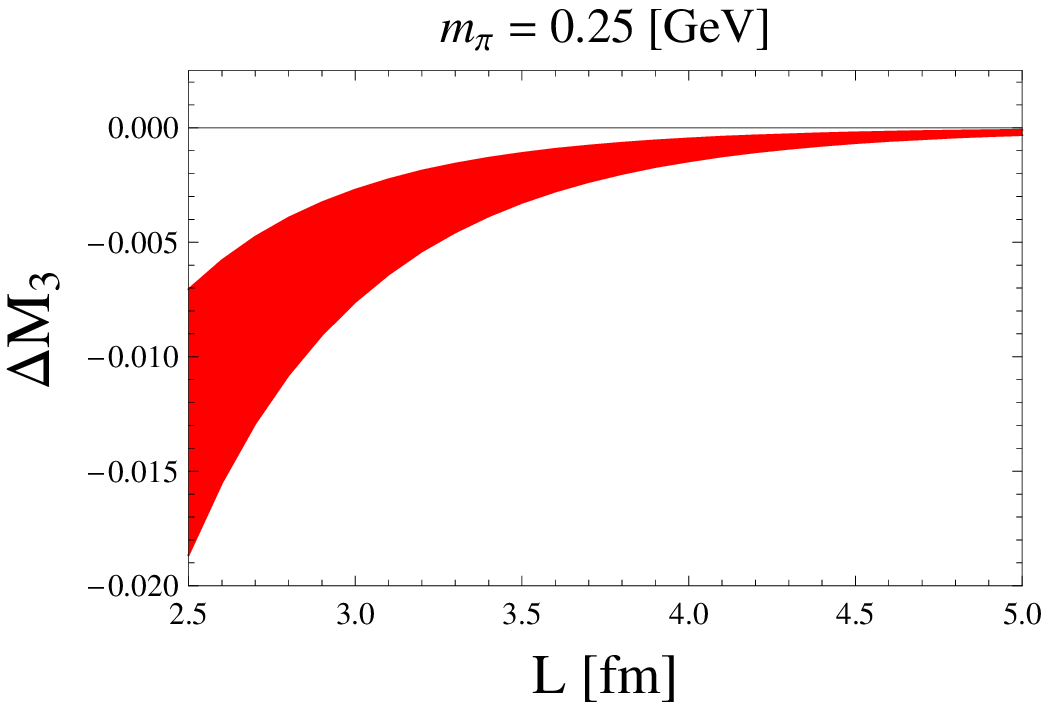,width=3in}
$\quad \quad$
\epsfig{file=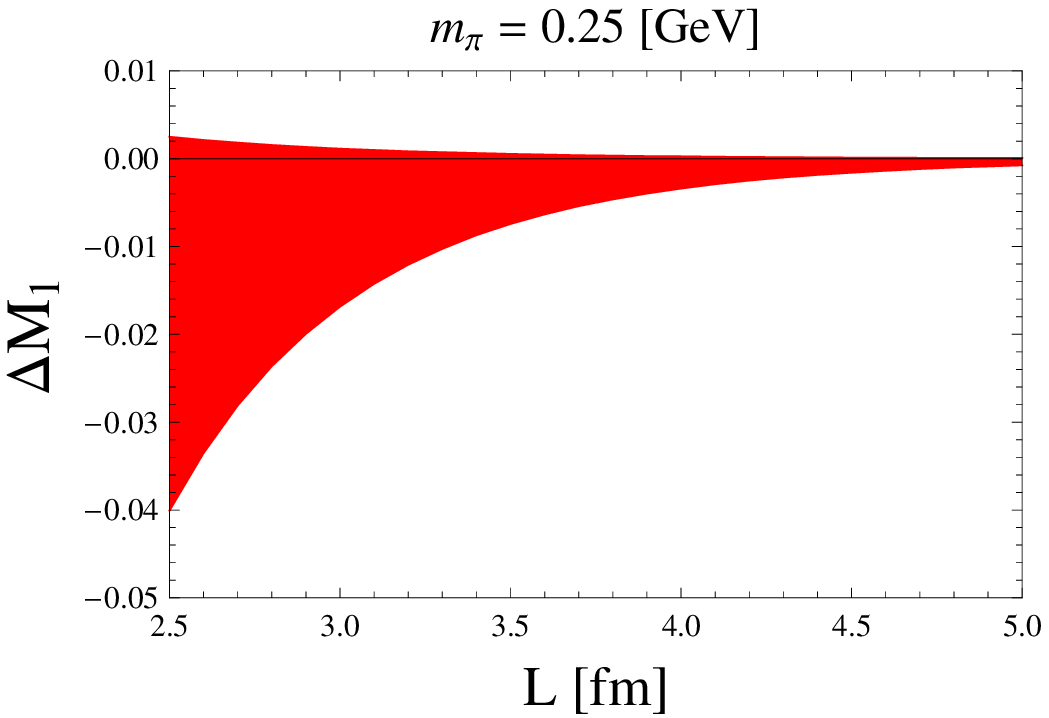,width=3in}
$\quad \quad$
\epsfig{file=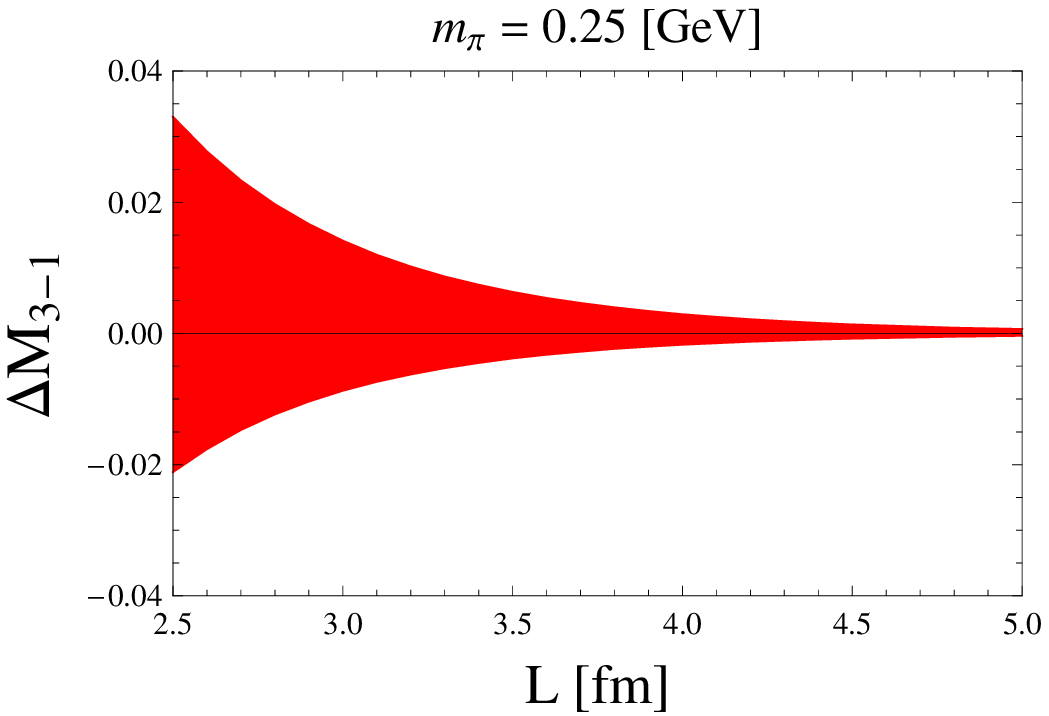,width=3in}
\caption{
Numerical estimates for the maximal nucleon splittings at finite volume, 
$\D M_{\bm{3}}$, $\D M_{\bm{1}}$, and $\D M_{\bm{3} - \bm{1}}$
given in Eqs.~\eqref{eq:3split}, \eqref{eq:1split}, and \eqref{eq:31split}, respectively. 
We plot each relative splitting as a function of  $L$ with the lattice pion mass
fixed at $m_\pi = 0.25 \, \texttt{GeV}$. 
The twist angles are fixed at $\bm{\theta} = \pi (1, 1, 1)$ to give the maximal splittings.
The bands arise from variation of the parameter $g_1$ assuming naturalness.
}
\label{f:IsoSplit}
\end{center}
\end{figure}

Numerically we can estimate the nucleon splittings
by using phenomenological input for the low-energy constants: 
$g_A = 1.25$, 
$g_{\D N} = 1.5$, 
$\D = 0.29 \, \texttt{GeV}$,
$M_N = 0.94 \, \texttt{GeV}$,
and 
$f = 0.13 \, \texttt{GeV}$. 
For the unknown partially quenched axial coupling 
$g_1$, 
we assume it is of natural size and vary it within the rage
$- 2 \leq g_1 \leq 2$. 
Each of the mass splittings $\D M_{\bm{3}}$, $\D M_{\bm{1}}$, 
and $\D M_{\bm{3} - \bm{1}}$ is a maximum when 
$\bm{\theta} = \pi ( 1, 1, 1)$. 
We choose this value for $\bm{\theta}$ to investigate the 
worst case scenario. 
In Figure~\ref{f:IsoSplit}, we investigate the relative mass splittings' 
dependence on 
$L$ 
for a fixed value of
$m_\pi$, which is chosen to be $0.25 \, \texttt{GeV}$.
Values shown for the maximal splittings are all less than five percent.
We will thus neglect the nucleon splittings in our analysis below.%
\footnote{ 
Additionally partially twisted isospin splittings in the meson sector
have been shown to be negligible on current sized lattices~\cite{Jiang:2006gna}. 
The same is true of the infrared renormalization of the twist angles.
These effects will hence also be neglected.
}

\section{Nucleon isovector form factors} \label{s:vec}

Electromagnetic form factors appear in vector current matrix elements of the nucleon. 
In terms of Dirac and Pauli form factors 
denoted by 
$F_1^N(\bm{Q}^2)$ 
and 
$F_2^N(\bm{Q}^2)$, 
respectively,
the nucleon current matrix element has the decomposition
\begin{equation} \label{eq:FFs}
\langle N (\bm{P}') | J^{\text{em}}_\mu | N (\bm{P}) \rangle
=
\ol u(\bm{P}') 
\left[
\gamma_\mu F^N_1(\bm{Q}^2)
-
\frac{\sigma_{\mu \nu} Q_\nu }{2 M_N}
F^N_2(\bm{Q}^2)
\right]
u(\bm{P})
,\end{equation}
where $Q_\mu = (P' - P)_\mu$ is the momentum transfer.
In QCD, the electromagnetic current is given as
$J^{\text{em}}_\mu = q_u \, \ol u \gamma_\mu u + q_d \, \ol d \gamma_\mu d$.
In heavy baryon chiral perturbation theory, 
the decomposition of the current appears in terms of the Sachs 
electric and magnetic form factors, 
$G_E^N(\bm{Q}^2)$ and $G_M^N(\bm{Q}^2)$, i.e.
one has
\begin{equation}
\langle N_v (\bm{P}') | J^{\text{em}}_\mu | N_v (\bm{P}) \rangle
=
\ol u_v
\left[
v_\mu G^N_E(\bm{Q}^2)
-
\frac{[S_\mu, S_\nu] Q_\nu }{M_N}
G^N_M(\bm{Q}^2)
\right]
u_v
\label{eq:Sachs}
,\end{equation}
with the relations
\begin{eqnarray}
G_{E}^N(\bm{Q}^2) &=& F_1^N(\bm{Q}^2)  + \frac{Q^2}{4 M_N^2} F_2^N(\bm{Q}^2)
\\
G_M^N(\bm{Q}^2) &=& F_1^N(\bm{Q}^2) + F^N_2(\bm{Q}^2)
.\end{eqnarray}
We have appended velocity subscripts in Eq.~\eqref{eq:Sachs} for clarity. 
The $u_v$ are two-component Pauli spinors.

To calculate these form factors with twisted boundary conditions on the lattice, 
one writes the current matrix element in terms of the various quark contractions
with the electromagnetic current. 
The propagators coupling to the current in each contraction we call the active quark propagators. 
These are the propagators determined with twisted boundary conditions. 
Omitted from this calculation are the current insertions on quark lines that are self-contracted.
These disconnected contributions are notoriously difficult to calculate using lattice QCD. 
This difficulty notwithstanding, their contributions cannot be modified to produce continuous momentum 
transfer between the initial and final state hadron. 
As with present-day lattice calculations, we too will omit these contributions but with the caveat
that their eventual inclusion will be limited to hadrons with Fourier momentum modes of the lattice.

Now we discuss precisely how to calculate the connected part of the nucleon form factors in the effective theory.
To specialize to the application of  twisted boundary conditions on the active quarks, 
we must separate the current into two pieces, 
\begin{eqnarray}
J_\mu^1 &=& q_u \, \ol u_2 \gamma_\mu u_1 \\
J_\mu^2 &=& q_d \, \ol d_2 \gamma_\mu d_1 
.\end{eqnarray}
By evaluating matrix elements of 
$J_\mu^1$ with $\bm{\theta}^u = \bm{\theta}$, $\bm{\theta'}^u = \bm{\theta'}$, 
and 
$J_\mu^2$ with $\bm{\theta}^d = \bm{\theta}$, $\bm{\theta'}^d = \bm{\theta'}$,
both currents induce momentum transfer from 
$\bm{P} = \bm{\theta}/ L$ 
to 
$\bm{P}' = \bm{\theta'}/L$. 
First let us consider proton matrix elements. 
Considering the quark-level contractions, we find
\begin{multline}
\frac{3}{2}
\big\langle N_{\bm{1}} (u u_2 d_1)  \big|  J_\mu^1 \big| N_{\bm{1}} (u u_1 d_1) \big\rangle
\Bigg|_{\bm{\theta}^d = \bm{0} }
+
\frac{1}{2}
\big\langle N_{\bm{3}} (u u_2 d_1)  \big|  J_\mu^1 \big| N_{\bm{3}} (u u_1 d_1) \big\rangle
\Bigg|_{\bm{\theta}^d = \bm{0} }
\\
+ 
\big\langle N_{\bm{3}} (u u d_2) \big| J_\mu^2 \big| N_{\bm{3}}(u u d_1) \big\rangle
\overset{L \to \infty}{\longrightarrow}
\langle p(\bm{P}') | J_\mu^{\text{em}} | p(\bm{P}) \rangle_{\text{connected}}
\label{eq:recipe}
.\end{multline}
The subscripts on $N$ refer to the representation under untwisted isospin, 
and parenthetically we list the quark content. 
We treat the active quark twists as implicit: each is from an initial quark, 
$u_1$ or $d_1$, 
with twist 
$\bm{\theta}$ 
to a final quark,
$u_2$ or $d_2$,  
with twist 
$\bm{\theta'}$. 
We stress that Eq.~\eqref{eq:recipe} provides the recipe for calculating in the effective theory
what is implemented on the lattice by twisting the active quarks.
For the neutron there is a similar construction, however, it is easiest
to appeal to charge symmetry (isospin rotation by $\pi/2$) from which follows the relation
\begin{equation}
\langle n(\bm{P}') | J_\mu^{\text{em}} | n(\bm{P}) \rangle
=
\langle p(\bm{P}') | J_\mu^{\text{em}} | p(\bm{P}) \rangle \Bigg|_{q_u \leftrightarrow q_d}
.\end{equation}

In partially quenched QCD, the current is defined by 
$J^{a}_ {\mu} = \ol Q \gamma_\mu \ol \cQ {}^a Q$. 
The choice of supermatrices 
$\ol \cQ {}^a$ 
used to extend the charges is not unique \cite{Golterman:2001qj}.
One should choose a form of the supermatrices that maintains the cancellation of valence and ghost quark loops
with an operator insertion~\cite{Tiburzi:2004mv}. 
For the flavor changing currents we consider, 
the simplest choice
$(\ol \cQ {}^1)_{ij} = q_u \,  \delta_{i3} \, \delta_{j2}$
and 
$(\ol \cQ {}^2)_{ij} = q_d \,  \delta_{i5} \, \delta_{j4}$
results in the correct physics. 
This is because operator self-contractions automatically vanish; thus, 
any non-zero charges in the ghost sector must ultimately yield zero,
and consequently would be superfluous.
Charges in the sea, while not superfluous, are absent due to
restricting to the connected part of three-point functions.

Operators that contribute at tree-level to the electromagnetic currents in $SU(2)$ 
chiral perturbation theory are contained in the Lagrangian
\begin{eqnarray}
\cL 
&=& 
-\frac{i  \mu_0}{2 M_N} \left( \ol N [ S_\mu, S_\nu]  N \right) \tr ( \cQ )  F_{\mu \nu}
-\frac{i  \mu_I}{2 M_N} \left( \ol N [ S_\mu, S_\nu] \cQ N \right)  F_{\mu \nu}
\notag \\
&& 
-\frac{c_0  }{\L_\chi^2}  \left( \ol N N \right) \tr ( \cQ ) v_\mu \partial_\nu F_{\mu \nu} 
-\frac{c_I  }{\L_\chi^2}  \left( \ol N \cQ N \right) v_\mu \partial_\nu F_{\mu \nu} 
\label{eq:current2}
,\end{eqnarray}
where $\cQ = \diag ( q_u, q_d )$ is the electric charge matrix. 
The operators with coefficients $\mu_0$ and $\mu_I$ give the leading local contributions to the magnetic moments, 
while the operators with coefficients $c_0$ and $c_I$ give the leading local contributions to the electric charge radii. 
Operators for the magnetic radii occur at one higher order than the leading loop contributions. 
The combination $\frac{2}{3} \mu_0 + \frac{1}{3} \mu_I$ is isoscalar, while $\mu_I$ is isovector. 
Analogous linear combinations of $c_0$ and $c_I$ form isoscalar and isovector contributions to the charge radius.
In writing down the analogous terms in the partially quenched chiral Lagrangian, 
$\cQ \to \ol \cQ {}^a$. 
Because of the condition $\str (\ol \cQ {}^a ) = 0$, 
we see that there will be missing information in the partially quenched theory:
there will only be operators where $\ol \cQ {}^a$ transforms under the adjoint,  
because of the lack of singlet component. 
This is the effective theory manifestation of neglecting the disconnected contributions. 
Consequently we will not be sensitive to the isoscalar combination low-energy constants.

To consider the baryon current
in partially twisted, partially quenched chiral perturbation theory, 
we promote $\ol \cQ {}^a$ 
from the specific form used in our calculations to the most general
form transforming under both the adjoint and singlet of $SU(7|5)$. 
The baryon current has the form
\begin{eqnarray}
\d J^{a}_{\mu} 
&=&
- \frac{i}{2 M_N} 
\left\{
\mu_\a  \hat{D}_\nu \left( \ol \cB [S_\mu, S_\nu] \cB \ol \cQ {}^a  \right)
+
\mu_\b  \hat{D}_\nu \left( \ol \cB [S_\mu, S_\nu] \ol \cQ {}^a \cB \right)
+
\mu_\gamma  \hat{D}_\nu \left( \ol \cB [S_\mu, S_\nu]  \cB \right) \, \str ( \ol \cQ {}^a )
\right\} \notag \\
&&  
- \frac{1}{\L_\chi^2} 
\Bigg\{ 
c_\a 
\left[ 
\hat{D}_\mu \hat{D}_\nu \left( \ol \cB v_\nu \cB \ol \cQ {}^a  \right)
- 
\hat{D}^2 \left( \ol \cB v_\nu \cB \ol \cQ {}^a  \right)
\right]
+c_\b 
\left[ 
\hat{D}_\mu \hat{D}_\nu \left( \ol \cB v_\nu \ol \cQ {}^a \cB   \right)
- 
\hat{D}^2 \left( \ol \cB v_\nu \ol \cQ {}^a \cB \right)
\right]
\notag \\
&&
\phantom{spacers}
+c_\gamma 
\left[ 
\hat{D}_\mu \hat{D}_\nu \left( \ol \cB v_\nu  \cB   \right)
- 
\hat{D}^2 \left( \ol \cB v_\nu \cB \right)
\right] \, \str ( \ol \cQ {}^a )
\Bigg\}
\label{eq:currentB}
.\end{eqnarray}
Restricting all quark indices in Eq.~\eqref{eq:currentB}
to the sea sector, we can match onto the  
nucleon current of two-flavor chiral perturbation theory
in Eq.~\eqref{eq:current2}. 
Matching with the physical light quark charges yields the relations
\begin{align}
& \mu_0 = \frac{1}{6} \mu_\a + \frac{2}{3} \mu_\b +  \mu_\gamma, 
&\mu_I = \frac{2}{3} \mu_\a - \frac{1}{3} \mu_\b, \\
& c_0 = \frac{1}{6} c_\a + \frac{2}{3} c_\b +  c_\gamma,  
&c_I = \frac{2}{3} c_\a - \frac{1}{3} c_\b
,\end{align}
between the partially quenched low-energy constants and
the physical parameters of chiral perturbation theory. 
Because our current lacks a flavor singlet component, 
$\str ( \ol \cQ {}^a ) = 0$, 
the constants $\mu_\gamma$ and $c_\gamma$ will 
always be absent from our expressions for nucleon current matrix elements. 
Consequently only the isovector combinations will be 
expressible in terms of physical parameters, specifically
$\mu_I$ and $c_I$. 
The isoscalar combinations will always contain unphysical low-energy 
constants even at the unitary mass point $m_j = m_u$.

\subsection{Infinite Volume}
\label{s:Infinite}

A useful check on our formulation of current matrix elements 
for partially twisted boundary conditions is the infinite volume limit. 
In this limit, we must recover the connected parts of the proton
and neutron form factors. 
These results, moreover, 
show the consequences of vanishing sea quark charges,
and are of use to lattice practitioners beyond the use of 
twisted boundary conditions.

\begin{figure}
\begin{center}
\epsfig{file=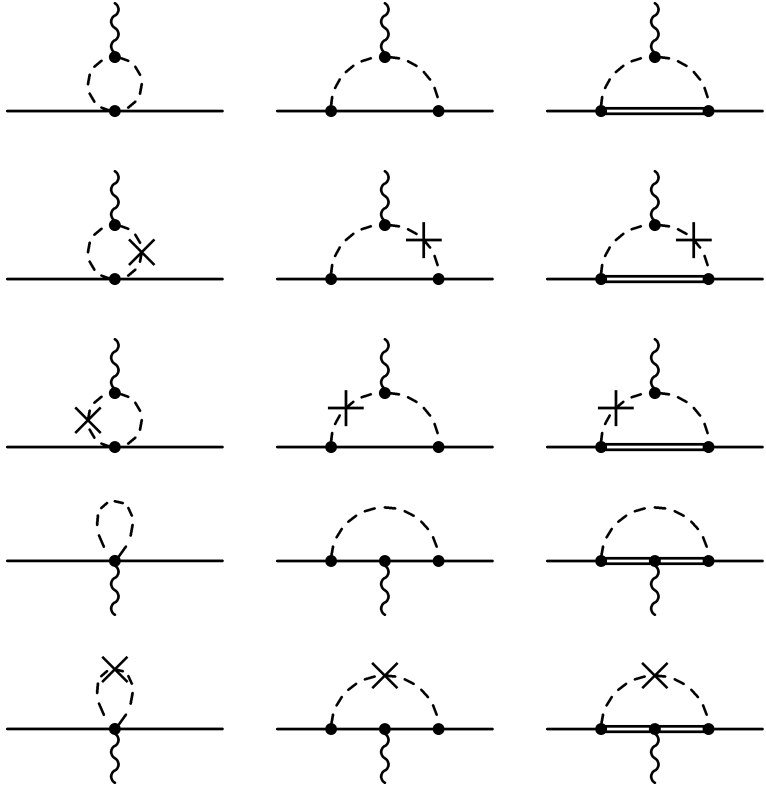,width=4in}
\caption{Diagrams contributing to the nucleon vector current in partially quenched chiral perturbation theory.
A thin (thick) line denotes a spin-$1/2$ (spin-$3/2$) baryon, while a dashed line denotes
a meson. Partially quenched hairpins are depicted by a crossed dashed line
and the wiggly line represents the vector current. }
\label{f:Nvec}
\end{center}
\end{figure}

The calculation of the current matrix elements in Eq.~\eqref{eq:recipe} 
can be split into two parts. 
There are local contributions and loop contributions. 
The local terms are easiest: there are Born level charge couplings
contained in the free Lagrangian~\eqref{eq:LB}, 
and there are additional local contributions from higher-order operators
appearing in the baryon current~\eqref{eq:currentB}. 
The loop contributions are generated from the pion-nucleon-nucleon
and pion-nucleon-delta interactions contained in the Lagrangian Eq.~\eqref{eq:Lint}. 
The relevant diagrams are depicted in Figure~\ref{f:Nvec}. 
Additionally at this order, we need to multiply the Born-level couplings
by the wavefunction renormalization which arises from the diagrams in 
Figure~\ref{f:Nmass}.

To express the form factors, we define 
the three momentum transfer
$\bm{Q} = \bm{q} + \bm{B}' - \bm{B}$, 
and the quantity 
\begin{equation}
P_\phi = \sqrt{1 + \frac{x(1-x) \bm{Q}^2}{m_\phi^2}}
.\end{equation}
Here we are additionally considering the 
nucleon with Fourier momentum transfer 
$\bm{q} = 2 \pi \bm{n} / L$, where $\bm{n}$ is 
a triplet of integers. 
Expressing the form factors in terms of $\bm{Q}$, 
we can easily generalize between the 
untwisted case $\bm{B} = \bm{B}' = \bm{0}$, 
and the case of zero lattice momentum $\bm{q} = \bm{0}$. 
In infinite volume, all results can be expressed as 
a function of $\bm{Q}$.

For the proton, the local and loop contributions 
produce the connected part of the electric form factor
\begin{eqnarray}
G_E^p(\bm{Q}^2) 
&=&  
2 q_u  +  q_d  
+ 
\frac{\bm{Q}^2}{ 6 \L_\chi^2} \left[ \  q_u  ( 5 c_\a + 2 c_\b) + q_d ( c_\a + 4 c_\b)  \right] 
\notag \\
&& + 
\frac{2}{(4 \pi f)^2}
\int_0^1 dx \, 
( 2 q_u + q_d)
\left\{
\frac{1}{6} \bm{Q}^2 \log \frac{m^2_{ju}}{\mu^2}
+
m_{ju}^2 P_{ju}^2 
\log P_{ju}^2 
\right\}
\notag \\
&& +
\frac{1}{(4 \pi f)^2} 
\int_0^1 dx \Bigg\{
\b_\pi
\left[
-\frac{5}{6} \bm{Q}^2 \log \frac{m_\pi^2}{\mu^2}
+  m_\pi^2 ( 2 - 5 P_\pi^2)  \log P_\pi^2
\right] 
\notag \\
&& \phantom{phantomspac} + 
\b_{ju}
\left[
-\frac{5}{6} \bm{Q}^2 \log \frac{m_{ju}^2}{\mu^2}
+  m_{ju}^2 ( 2 - 5 P_{ju}^2) \log P_{ju}^2
\right] 
\Bigg\}
\notag \\
&& +
\frac{6 g_{\D N}^2}{(4 \pi f)^2} \int_0^1 dx
\Bigg\{ 
\b'_\pi 
\left[ J(m_\pi P_\pi, \D, \mu) - J(m_\pi, \D, \mu) 
+ 
\frac{2}{3} x(1-x) \bm{Q}^2 G(m_\pi P_\pi, \D) 
\right]
\notag \\
&& \phantom{space} +
\b'_{ju} 
\left[
J(m_{ju} P_{ju}, \D, \mu) - J(m_{ju}, \D, \mu) 
+ 
\frac{2}{3} x(1-x) \bm{Q}^2 G(m_{ju} P_{ju}, \D) 
\right]
\Bigg\}.
\notag \\
\end{eqnarray}
The coefficients from contributing loop mesons are given by
\begin{eqnarray}
\b_\pi &=& - \frac{1}{3} ( g_A^2 - g_A g_1 + g_1^2 / 4)  (q_u - q_d), \notag \\
\b_{ju} &=& - \frac{1}{3} \left( 4 g_A^2 + 2 g_A g_1 + g_1^2 \right) q_u - \frac{1}{2}  g_1^2 q_d, 
\label{eq:beta}
\end{eqnarray}
for loops containing spin-$1/2$ intermediate state baryons, and
\begin{eqnarray}
\b'_\pi &=& - \frac{1}{6} (q_u - q_d), \notag  \\
\b'_{ju} &=& \frac{1}{9} ( q_u + 2 q_d), 
\label{eq:betaprime}
\end{eqnarray}
for loop containing spin-$3/2$ intermediate state baryons. 
From the pion coefficients, one can clearly see the photon's coupling to the total 
charge of the pion, $q_u - q_d$. 
The valence-sea meson coefficients, 
however, 
reflect that the photon couples to only the valence quarks. 
The connected part of the proton magnetic form factor is given by
\begin{eqnarray}
G_M^p(\bm{Q}^2) 
&=& 
\frac{1}{6} \left[ \  q_u  ( 5 \mu_\a + 2 \mu_\b) + q_d ( \mu_\a + 4 \mu_\b)  \right]  
+
\frac{M_B}{4 \pi f^2} \int_0^1 dx 
\Bigg[  
\b _\pi m_\pi  P_\pi 
+
\b_{ju} m_{ju} P_{ju} 
\Bigg]
\notag \\
&& 
+ 
\frac{M_B g_{\D N}^2 }{4 \pi^2 f^2} \int_0^1 dx  
\left[
\b'_\pi  F(m_\pi P_\pi, \D) 
+
\b'_{ju} F(m_{ju} P_{ju}, \D)  
\right].
\end{eqnarray}
In writing the above expressions we have made use of abreviations for the non-analytic functions 
encountered from loop graphs. 
These functions are
\begin{eqnarray}
F(m,\d) 
&=& 
- \d \log \frac{m^2}{4 \d^2} 
+ 
\sqrt {\d^2 - m^2} \log \frac{\d - \sqrt{\d^2 - m^2 + i \varepsilon}}{\d + \sqrt{\d^2 - m^2 + i \varepsilon}},
\\
G(m,\d)  
&=&
\log \frac{m^2}{4 \d^2}
- \frac{\d}{\sqrt{\d^2 - m^2}}
\log \frac{\d - \sqrt{\d^2 - m^2 + i \varepsilon}}{\d + \sqrt{\d^2 - m^2 + i \varepsilon}},
\\
J(m,\d,\mu) 
&=&
m^2  \log \frac{m^2}{\mu^2}
- 
2 \d^2 \log \frac{m^2}{4 \d^2}
+ 
2 \d \sqrt{\d^2 - m^2}
\log \frac{\d - \sqrt{\d^2 - m^2 + i \varepsilon}}{\d + \sqrt{\d^2 - m^2 + i \varepsilon}}
,\end{eqnarray}
and have been renormalized to vanish in the chiral limit. 
The neutron electric and magnetic form factors can be deduced 
from the above expressions by swapping the electric charges
\begin{equation}
G_E^n(\bm{Q}^2) = G_E^p(\bm{Q}^2) \Bigg|_{q_u \leftrightarrow q_d}, 
\quad
\text{and}
\quad
G_M^n(\bm{Q}^2) = G_M^p(\bm{Q}^2) \Bigg|_{q_u \leftrightarrow q_d}
.\end{equation}

Let us focus first on the connected proton
form factors at the unitary mass point $m_j = m_u$.
Using the physical valence quark charges, we
have the connected proton electric form factor
\begin{eqnarray}
G_E^p(\bm{Q}^2) 
&=&  
1
+ 
\frac{ c_\a \bm{Q}^2}{ 2 \L_\chi^2} 
+
\frac{2}{(4 \pi f)^2}
\int_0^1 dx \, 
\left\{
\frac{1}{6} \bm{Q}^2 \log \frac{m^2_{\pi}}{\mu^2}
+
m_{\pi}^2 P_{\pi}^2 
\log P_{\pi}^2 
\right\}
\notag \\
&& -
\frac{1}{9 (4 \pi f)^2}  
( 11 g_A^2 + g_A g_1 + 5 g_1^2 / 4 )
\int_0^1 dx 
\left[
-\frac{5}{6} \bm{Q}^2 \log \frac{m_\pi^2}{\mu^2}
+  m_\pi^2 ( 2 - 5 P_\pi^2)  \log P_\pi^2
\right] 
\notag \\
&& -
\frac{g_{\D N}^2}{(4 \pi f)^2} \int_0^1 dx 
\left[ J(m_\pi P_\pi, \D, \mu) - J(m_\pi, \D, \mu) 
+ 
\frac{2}{3} x(1-x) \bm{Q}^2 G(m_\pi P_\pi, \D) 
\right].
\notag \\
\end{eqnarray}
Compared to full electric form factor, the connected contribution
has the wrong coefficients for the tadpole and delta loop contributions. 
It depends, moreover, on the unphysical low-energy constants 
$c_\a$ and $g_1$,
which survive as artifacts of quenching the sea quark charges.
The situation is similar with respect to the connected contribution
to the proton magnetic form factor
\begin{eqnarray}
G_M^p(\bm{Q}^2) 
&=& 
\frac{1}{2} \mu_\a
-
\frac{M_B}{36 \pi f^2} 
( 11 g_A^2 + g_A g_1 + 5 g_1^2 / 4 )
\int_0^1 dx  \,
m_\pi  P_\pi 
- 
\frac{M_B g_{\D N}^2 }{24 \pi^2 f^2} \int_0^1 dx  \,
F(m_\pi P_\pi, \D).
\notag \\
\end{eqnarray}
Compared to the full magnetic form factor
the delta contribution does not have the correct numerical factor, and
the result depends on unphysical parameters $\mu_\a$ and $g_1$.

Connected neutron form factors suffer analogous maladies as the reader can easily verify. 
By contrast, the isovector form factors have the correct form. 
These form factors are defined as the difference between proton and neutron 
form factors 
\begin{eqnarray}
G_E^v(\bm{Q}^2) &=& G_E^p(\bm{Q}^2) - G_E^n(\bm{Q}^2), \\
G_M^v(\bm{Q}^2) &=& G_M^p(\bm{Q}^2) - G_M^n(\bm{Q}^2) 
.\end{eqnarray}
In the isospin limit, 
the disconnected operator insertion must cancel out of the isovector combinations. 
Using the connected form factors for the proton and neutron, we find
\begin{eqnarray}
G_E^v(\bm{Q}^2) 
&=&  
1
+ 
c_I
\frac{\bm{Q}^2}{ \L_\chi^2} 
+ 
\frac{2}{(4 \pi f)^2}
\int_0^1 dx \, 
\left\{
\frac{1}{6} \bm{Q}^2 \log \frac{m^2_{ju}}{\mu^2}
+
m_{ju}^2 P_{ju}^2 
\log P_{ju}^2 
\right\}
\notag \\
&& -
\frac{1}{6 (4 \pi f)^2} 
\int_0^1 dx \Bigg\{
g_{\pi N N}^2
\left[
-\frac{5}{6} \bm{Q}^2 \log \frac{m_\pi^2}{\mu^2}
+  m_\pi^2 ( 2 - 5 P_\pi^2)  \log P_\pi^2
\right] 
\notag \\
&& \phantom{phantomspac} + 
g_{ju N N}^2
\left[
-\frac{5}{6} \bm{Q}^2 \log \frac{m_{ju}^2}{\mu^2}
+  m_{ju}^2 ( 2 - 5 P_{ju}^2) \log P_{ju}^2
\right] 
\Bigg\}
\notag \\
&& -
\frac{2 g_{\D N}^2}{(4 \pi f)^2} \int_0^1 dx
\Bigg\{ 
J(m_\pi P_\pi, \D, \mu) - J(m_\pi, \D, \mu) 
+ 
\frac{2}{3} x(1-x) \bm{Q}^2 G(m_\pi P_\pi, \D) 
\notag \\
&& \phantom{space} +
\frac{1}{3} 
\left[
J(m_{ju} P_{ju}, \D, \mu) - J(m_{ju}, \D, \mu) 
+ 
\frac{2}{3} x(1-x) \bm{Q}^2 G(m_{ju} P_{ju}, \D) 
\right]
\Bigg\},
\notag \\
\end{eqnarray}
where we have abbreviated the combination of couplings 
\begin{eqnarray}
g_{\pi NN}^2 &=& 4 g_A^2 - 4 g_A g_1 + g_1^2, \notag \\ 
g_{ju NN}^2 &=& 8 g_A^2 + 4 g_A g_1 - g_1^2.
\label{eq:PQcouplings}
\end{eqnarray} 
These appear as effective axial couplings squared 
for pion and valence-sea meson loops in partially quenched chiral perturbation theory after summing over degenerate mesons. 
The partially quenched isovector magnetic form factor is
\begin{eqnarray}
G_M^v(\bm{Q}^2) 
&=& 
\mu_I 
-
\frac{M_B}{24 \pi f^2} \int_0^1 dx 
\Bigg[  
g_{\pi N N}^2  m_\pi  P_\pi 
+
g_{ju N N}^2  m_{ju} P_{ju} 
\Bigg]
\notag \\
&& 
- 
\frac{M_B g_{\D N}^2 }{12 \pi^2 f^2} \int_0^1 dx  
\left[
F(m_\pi P_\pi, \D) 
+
\frac{1}{3} F(m_{ju} P_{ju}, \D)  
\right].
\end{eqnarray}
These partially quenched form factors agree with those determined in 
$SU(4|2)$ partially quenched chiral perturbation theory~\cite{Beane:2002vq,Arndt:2003ww}.
The local contributions are now proportional to  $\mu_I$ and $c_I$, which are physical parameters. 
Both of these partially quenched form factors, however, depend on the unphysical coupling
$g_1$. 
Taking the valence-sea meson to be degenerate with the pion,
$m_{ju}^2 = m_\pi^2$, 
this dependence disappears because 
$g_{\pi NN}^2 + g_{ju NN}^2 = 12 g_A^2$. 
It is only in this limit that the isovector form factors reproduce the correct QCD physics.%
\footnote{ 
This point is often overlooked, 
particularly in mixed action simulations
which are automatically partially quenched. 
In a mixed action simulation, the valence and sea pion masses
are tuned in order to mitigate unitarity violations. 
The valence-sea meson mass,
however, 
is not protected from additive renormalization and
is degenerate with the pion only in the strict continuum limit.
} 
For completeness, 
the nucleon isovector form factors 
resulting from taking   
$m_{ju}^2 = m_\pi^2$
are 
\begin{eqnarray}
G_E^v(\bm{Q}^2) 
&=&  
1
+ 
c_I
\frac{\bm{Q}^2}{ \L_\chi^2} 
+ 
\frac{2}{(4 \pi f)^2}
\int_0^1 dx \, 
\left[
\frac{1}{6} \bm{Q}^2 \log \frac{m^2_{\pi}}{\mu^2}
+
m_{\pi}^2 P_{\pi}^2 
\log P_{\pi}^2 
\right]
\notag \\
&& -
\frac{2 g_A^2}{ (4 \pi f)^2} 
\int_0^1 dx 
\left[
-\frac{5}{6} \bm{Q}^2 \log \frac{m_\pi^2}{\mu^2}
+  m_\pi^2 ( 2 - 5 P_\pi^2)  \log P_\pi^2
\right] 
\notag \\
&& -
\frac{8 g_{\D N}^2}{3 (4 \pi f)^2} \int_0^1 dx
\Bigg[ 
J(m_\pi P_\pi, \D, \mu) - J(m_\pi, \D, \mu) 
+ 
\frac{2}{3} x(1-x) \bm{Q}^2 G(m_\pi P_\pi, \D) 
\Bigg],
\notag \\
\label{eq:GEv}
\end{eqnarray}
for the isovector electric, and
\begin{eqnarray} \label{eq:GMv}
G_M^v(\bm{Q}^2) 
&=& 
\mu_I 
-
\frac{g_A^2 M_B}{2 \pi f^2} \int_0^1 dx 
\, 
m_\pi  P_\pi 
- 
\frac{M_B g_{\D N}^2 }{9 \pi^2 f^2} \int_0^1 dx  
\,
F(m_\pi P_\pi, \D) 
,\end{eqnarray}
for the isovector magnetic form factor. 
These results agree with the standard two-flavor chiral perturbation theory 
calculations in the literature~\cite{Bernard:1992qa,Bernard:1998gv}.

\subsection{Finite Volume}

\begin{figure}
\begin{center}
\epsfig{file=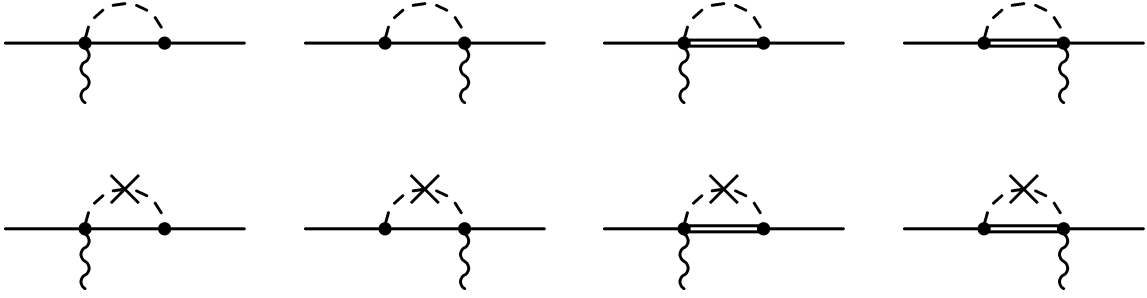,width=6in}
\caption{Additional diagrams contributing to the nucleon vector current in finite volume \PQCPT.
Diagram elements are the same as those in Figure~\ref{f:Nvec}.
In the infinite volume limit, these diagrams vanish by Lorentz invariance. 
}
\label{f:moreNvec}
\end{center}
\end{figure}

We now evaluate the matrix elements contributing to the connected 
part of the proton current in Eq.~\eqref{eq:recipe} in finite volume. 
This requires us to revisit the computation of the wavefunction 
renormalization diagrams shown in Figure~\ref{f:Nmass}, 
and the form factor diagrams shown in Figure~\ref{f:Nvec}. 
Additionally there are new contributing diagrams which are displayed in Figure~\ref{f:moreNvec}. 
These diagrams ordinarily vanish in infinite volume by Lorentz invariance. 
Furthermore at finite volume with periodic boundary conditions, 
these diagrams also vanish but by the remnant discrete rotational symmetry
(cubic invariance). 
With continuous twist angles, however, these diagrams do not vanish and are required
in our computation of current matrix elements.

Resulting expressions for the temporal and spatial components of the current are 
quite lengthy and are displayed in their entirety in Appendix~\ref{A:result}. 
For ease, the expressions given in this section will employ various simplifications. 
Firstly we work at the unitary mass point, $m_{ju}^2 = m_\pi^2$. 
We will focus on the connected proton result, 
as well as the isovector combination of finite volume matrix elements. 
Additionally as Lorentz symmetry is not respected at finite volume, 
the form factor decomposition in infinite volume is no longer valid, 
see~\cite{Tiburzi:2007ep}, for example.
With twisted boundary conditions, we find the temporal component
of the current acquires spin dependence at finite volume. 
Similarly the spatial components of the current acquire spin diagonal 
terms. These terms are displayed in Appendix~\ref{A:result}, 
while the expressions
presented here will either be unpolarized for the temporal component, 
or the difference of polarized matrix elements in the case of the spatial components. 
Lastly the results in Appendix~\ref{A:result} are for a general frame of reference 
in which the initial state moves with momentum $\bm{\theta}/L$, 
and the final state moves with $\bm{\theta'}/L$. 
The expressions given in this section will be specific to either
the rest frame, in which $\bm{\theta}=\bm{0}$, or the Breit frame, 
in which $\bm{\theta} = - \bm{\theta'}$.

\subsubsection{Rest Frame}

In the rest frame, 
the momentum transfer is given by $\bm{Q} = \bm{q} + \bm{B}'$.
The finite volume modifications to proton current matrix elements are given by
\begin{eqnarray}
\frac{1}{2} \sum_{m = \pm} 
\langle p, m |  \d J_4 | p, m \rangle
&=&  
\frac{1}{f^2}
\int_0^1 dx 
\left[ 
\cI_{1/2} (m_{\pi} P_{\pi}, x \bm{Q})
- 
\frac{1}{2} \cI_{1/2} (m_{\pi},\bm{0})
- 
\frac{1}{2} \cI_{1/2} (m_{\pi},\bm{B}')
\right] 
\notag \\
&& - 
\frac{1}{12  f^2}
\Bigg\{ 
( 11 g_A^2 +  g_A g_1 + 5 g_1^2 /4)
\left[
\ol  \cJ(m_\pi, \bm{0},0) 
+
\ol \cJ(m_\pi, \bm{B}', 0) \Big)
\right]
\notag \\
&& \phantom{spaspa}
-
3
g_{\D N}^2
\left[ 
\ol \cJ(m_\pi, \bm{0},\D) 
+
\ol \cJ(m_\pi, \bm{B}',\D) 
\right]
\Bigg\}
\notag \\
&& 
+
\frac{1}{6 f^2}
\int_0^1 dx 
\Big[
( 11 g_A^2 + g_A g_1 + 5 g_1^2 / 4 ) 
\cJ(m_{\pi} P_{\pi}, \bm{0}, \bm{Q}, x \bm{Q} , 0)
\notag \\
&& \phantom{space}
-
3 g_{\D N}^2 
\cJ(m_{\pi} P_{\pi}, \bm{0}, \bm{Q}, x \bm{Q}, \D)
\Big]
,\end{eqnarray}
for the unpolarized time component of the current; 
and, 
\begin{eqnarray}
\langle p , \pm | \d J_{i} | p, \mp \rangle
&=& 
\frac{1}{f^2} 
\langle \pm | \, [S_{k}, S_j] \, | \mp \rangle
\Bigg\{
\delta_{ki} 
\frac{1}{9}
( 13 g_A^2 + 11 g_A g_1 + g_1^2 / 4 )
\cK^j(m_\pi, \bm{B}', 0)
\notag \\
&& \phantom{space}
-
\delta_{ki}
\frac{ g_{\D N}^2 }{6}
\cK^j(m_\pi, \bm{B}', \D)
\notag \\
&& + 
\frac{1}{6} 
\bm{Q}_k
\int_0^1 dx 
\Big[
(11 g_A^2 + g_A g_1 + 5 g_1^2 / 4 ) 
\cL^{ji} (m_\pi P_\pi, \bm{0}, \bm{Q},  x \bm{Q}, 0) 
\notag \\ 
&& \phantom{space}
+
\frac{3}{2} g_{\D N}^2
\cL^{ji} (m_\pi P_\pi, \bm{0}, \bm{Q}, x \bm{Q}, \D)
\Big]
\Bigg\}
,\end{eqnarray}
for the spatial components.
We have chosen spin-flip matrix elements; 
these are simply related to differences of spin 
polarized matrix elements.

The finite volume corrections to isovector matrix elements, we write out similarly. 
\begin{eqnarray}
\frac{1}{2} \sum_{m = \pm} 
\langle p, m |  \d J^+_4 | n, m \rangle
&=&  
\frac{1}{f^2}
\int_0^1 dx 
\left[ 
\cI_{1/2} (m_{\pi} P_{\pi}, x \bm{Q})
- 
\frac{1}{2} \cI_{1/2} (m_{\pi},\bm{0})
- 
\frac{1}{2} \cI_{1/2} (m_{\pi},\bm{B}')
\right] 
\notag \\
&& - 
\frac{3}{2 f^2}
\Bigg\{
g_A^2 
\left[
\ol  \cJ(m_\pi, \bm{0},0) 
+
\ol \cJ(m_\pi, \bm{B}', 0) \Big)
\right]
\notag \\
&& \phantom{space}
-
\frac{4}{9}
g_{\D N}^2
\left[ 
\ol \cJ(m_\pi, \bm{0},\D) 
+
\ol \cJ(m_\pi, \bm{B}',\D) 
\right]
\Bigg\}
\notag \\
&& 
+
\frac{3}{f^2}
\int_0^1 dx 
\Big[
g_A^2 
\cJ(m_{\pi} P_{\pi}, \bm{0}, \bm{Q}, x \bm{Q} , 0)
- 
\frac{4}{9} 
g_{\D N}^2 
\cJ(m_{\pi} P_{\pi}, \bm{0}, \bm{Q}, x \bm{Q}, \D)
\Big]
, \notag \\
\label{eq:GEvRest}
\end{eqnarray}
for the unpolarized time component of the current; 
and, 
\begin{eqnarray}
\langle p , \pm | \d J^+_{i} | n, \mp \rangle
&=& 
\frac{1}{f^2} 
\langle \pm | \, [S_{k}, S_j] \, | \mp \rangle
\Bigg\{
2 \delta_{ki} 
( g_A^2 +  g_A g_1 )
\cK^j(m_\pi, \bm{B}', 0)
\notag \\
&& + 
3 
\bm{Q}_k
\int_0^1 dx 
\left[
g_A^2
\cL^{ji} (m_\pi P_\pi, \bm{0}, \bm{Q},  x \bm{Q}, 0) 
+
\frac{2}{9} g_{\D N}^2
\cL^{ji} (m_\pi P_\pi, \bm{0}, \bm{Q}, x \bm{Q}, \D)
\right]
\Bigg\}
,\notag \\
\label{eq:GMvRest}
\end{eqnarray}
for the spin-flip spatial current.

\subsubsection{Breit Frame}

In the Breit frame, we choose 
$\bm{B}' = - \bm{B}$ 
and the the momentum transfer is thus given by 
$\bm{Q} = \bm{q} - 2 \bm{B}$.
The finite volume modifications to proton current matrix elements are given by
\begin{eqnarray}
\frac{1}{2} \sum_{m = \pm} 
\langle p, m |  \d J_4 | p, m \rangle
&=&  
\frac{1}{f^2}
\int_0^1 dx 
\left[ 
\cI_{1/2} (m_{\pi} P_{\pi}, x \bm{Q}+ \bm{B})
- 
\cI_{1/2} (m_{\pi},\bm{B})
\right] 
\notag \\
&& - 
\frac{1}{6 f^2}
\Big[
( 11 g_A^2 +  g_A g_1 + 5 g_1^2 /4)
\ol \cJ(m_\pi, \bm{B}, 0) 
-
3 g_{\D N}^2
\ol \cJ(m_\pi, \bm{B},\D) 
\Big]
\notag \\
&& 
+
\frac{1}{6 f^2}
\int_0^1 dx 
\Big[
( 11 g_A^2 + g_A g_1 + 5 g_1^2 / 4 ) 
\cJ(m_{\pi} P_{\pi}, \bm{B}, \bm{Q} + \bm{B}, x \bm{Q}  + \bm{B}, 0)
\notag \\
&& \phantom{spaspaspa}
-
3 g_{\D N}^2 
\cJ(m_{\pi} P_{\pi}, \bm{B}, \bm{Q} + \bm{B}, x \bm{Q} + \bm{B}, \D)
\Big]
,\end{eqnarray}
for the unpolarized time component of the current; 
and, 
\begin{eqnarray}
\langle p , \pm | \d J_{i} | p, \mp \rangle
&=& 
\frac{1}{f^2} 
\langle \pm | \, [S_{k}, S_j] \, | \mp \rangle
\Bigg\{
- \delta_{ki} 
\frac{2}{9}
( 13 g_A^2 + 11 g_A g_1 + g_1^2 / 4 )
\cK^j(m_\pi, \bm{B}, 0)
\notag \\
&& 
\phantom{spacespacespace}
+
\delta_{ki}
\frac{ g_{\D N}^2 }{3}
\cK^j(m_\pi, \bm{B}, \D)
\notag \\
&& + 
\frac{1}{6} 
\bm{Q}_k
\int_0^1 dx 
\Big[
(11 g_A^2 + g_A g_1 + 5 g_1^2 / 4 ) 
\cL^{ji} (m_\pi P_\pi, \bm{B}, \bm{Q} + \bm{B},  x \bm{Q} + \bm{B}, 0) 
\notag \\
&& \phantom{spaspaspa}
+
\frac{3}{2} g_{\D N}^2
\cL^{ji} (m_\pi P_\pi, \bm{B}, \bm{Q} + \bm{B}, x \bm{Q} + \bm{B}, \D)
\Big]
\Bigg\}
,\end{eqnarray}
for the spatial components.

The isovector current matrix elements can similarly be derived in the Breit frame. 
For the time component of the current, we have
\begin{eqnarray}
\frac{1}{2} \sum_{m = \pm} 
\langle p, m |  \d J^+_4 | n, m \rangle
&=&  
\frac{1}{f^2}
\int_0^1 dx 
\left[ 
\cI_{1/2} (m_{\pi} P_{\pi}, x \bm{Q}+ \bm{B})
- 
\cI_{1/2} (m_{\pi},\bm{B})
\right] 
\notag \\
&& - 
\frac{3}{f^2}
\Big[
g_A^2 
\ol \cJ(m_\pi, \bm{B}, 0) 
-
\frac{4}{9} g_{\D N}^2
\ol \cJ(m_\pi, \bm{B},\D) 
\Big]
\notag \\
&& 
+
\frac{3}{f^2}
\int_0^1 dx 
\Big[
g_A^2
\cJ(m_{\pi} P_{\pi}, \bm{B}, \bm{Q} + \bm{B}, x \bm{Q}  + \bm{B}, 0)
\notag \\
&& \phantom{spaspaspa}
-
\frac{4}{9} g_{\D N}^2 
\cJ(m_{\pi} P_{\pi}, \bm{B}, \bm{Q} + \bm{B}, x \bm{Q} + \bm{B}, \D)
\Big]
\label{eq:GEvBreit}
.\end{eqnarray}
Finally, the spatial isovector current has the spin-flip finite volume 
corrections given by
\begin{eqnarray}
\langle p , \pm | \d J^+_{i} | n, \mp \rangle
&=& 
\frac{1}{f^2} 
\langle \pm | \, [S_{k}, S_j] \, | \mp \rangle
\Bigg\{
- 4 \delta_{ki} 
( g_A^2 + g_A g_1)
\cK^j(m_\pi, \bm{B}, 0)
\notag \\
&& + 
3 \bm{Q}_k
\int_0^1 dx 
\Big[
g_A^2
\cL^{ji} (m_\pi P_\pi, \bm{B}, \bm{Q} + \bm{B},  x \bm{Q} + \bm{B}, 0) 
\notag \\
&& \phantom{spaspaspa}
+
\frac{2}{9} g_{\D N}^2
\cL^{ji} (m_\pi P_\pi, \bm{B}, \bm{Q} + \bm{B}, x \bm{Q} + \bm{B}, \D)
\Big]
\Bigg\}
\label{eq:GMvBreit}
.\end{eqnarray}

\subsection{Numerical Estimates}

To estimate the effect of finite volume corrections, 
we use phenomenological input for the various coupling 
constants. The values we use have been listed
above in Section~\ref{s:mass}. 
We restrict our attention to isovector quantities
and numerically evaluate the corrections in the rest frame. 
We will comment on the qualitative behavior of volume
corrections in Breit frame.

\begin{figure}
\begin{center}
\epsfig{file=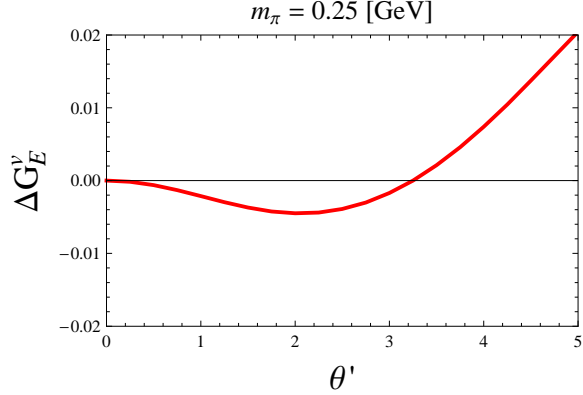,width=3.in}
\caption{
Relative change in the isovector electric form factor due to twisted boundary conditions in the rest frame.
Plotted versus the twisting angle $\theta'$ is $\D G_E^v(\bm{Q}^2,L)$ given in Eq.~\eqref{eq:GErel}.
The lattice size $L$ is fixed at $2.75 \, \texttt{fm}$,
and the momentum transfer is 
$|\bm{Q}| = \theta' / L =  \theta' \times 0.072 \, \texttt{GeV}$.
}
\label{f:dGE}
\end{center}
\end{figure}

Consider first the finite volume corrections to the
isovector electric form factor $G_E^v(\bm{Q}^2)$.%
\footnote{
Strictly speaking there are no longer 
electric and magnetic form factors on a torus
as the decomposition in Eq.~\eqref{eq:FFs}
relies on Lorentz invariance. 
We will use electric (magnetic) to denote 
quantities calculated from the temporal (spatial) 
component of the current with the appropriate spin structure. 
}
To access this form factor, we use unpolarized 
matrix elements of the time component of the current. 
We find
\begin{equation}
G_E^v(\bm{Q}^2, L) = G^v_E(\bm{Q}^2) + \d_L [ G_E^v(\bm{Q}^2) ]
,\end{equation}
where $G_E^v(\bm{Q}^2)$ is the infinite volume form factor
given by Eq.~\eqref{eq:GEv}, and $\d_L [G_E^v(\bm{Q}^2)]$ 
is the finite volume correction which is identical to the 
unpolarized matrix element in Eq.~\eqref{eq:GEvRest}. 
Here we work with the momentum transfer entirely due to twisting
$\bm{Q} = \bm{B}' = \bm{\theta'}/L$, 
and take $\bm{\theta'}$ to lie along one spatial direction. 
Notice the finite volume correction to the isovector electric form
factor is independent of any unphysical parameters, 
in particular the coupling $g_1$. 
The infinite volume isovector electric form factor
depends on the parameter $c_I(\mu)$, the value of 
which can be inferred from the charge radii of the 
proton and neutron. Using the Particle Data Group
averages~\cite{Amsler:2008zz}, we find
$c_I(\mu = 1 \, \texttt{GeV}) = -0.393$. 
In Figure~\ref{f:dGE}, we plot the relative change in the 
isovector electric form factor due to volume effects
$\D G^v_E$ defined by
\begin{equation} \label{eq:GErel}
\D G^v_E(\bm{Q}^2, L) =  \frac{G_E^v(\bm{Q}^2, L)  - G^v_E(\bm{Q}^2)}{G^v_E(\bm{Q}^2)}
.\end{equation}
Here we keep the box size fixed at $2.75 \, \texttt{fm}$, 
and plot versus the twisting angle $\theta'$.
Qualitatively the finite volume effect oscillates about 
the infinite volume form factor as $\theta'$ is increased.
The oscillations are damped, but this behavior is apparent 
at momentum transfers too large to trust the effective theory. 
There are no finite size effects at $\theta' = 0$ 
because of charge non-renormalization (which holds
due to treating the time direction as infinite). 
The results are shown for $m_\pi = 0.25 \, \texttt{GeV}$
with the finite volume effect generally at the percent level or less. 
The effect of finite volume is of course smaller for larger
pion masses.

\begin{figure}
\begin{center}
\epsfig{file=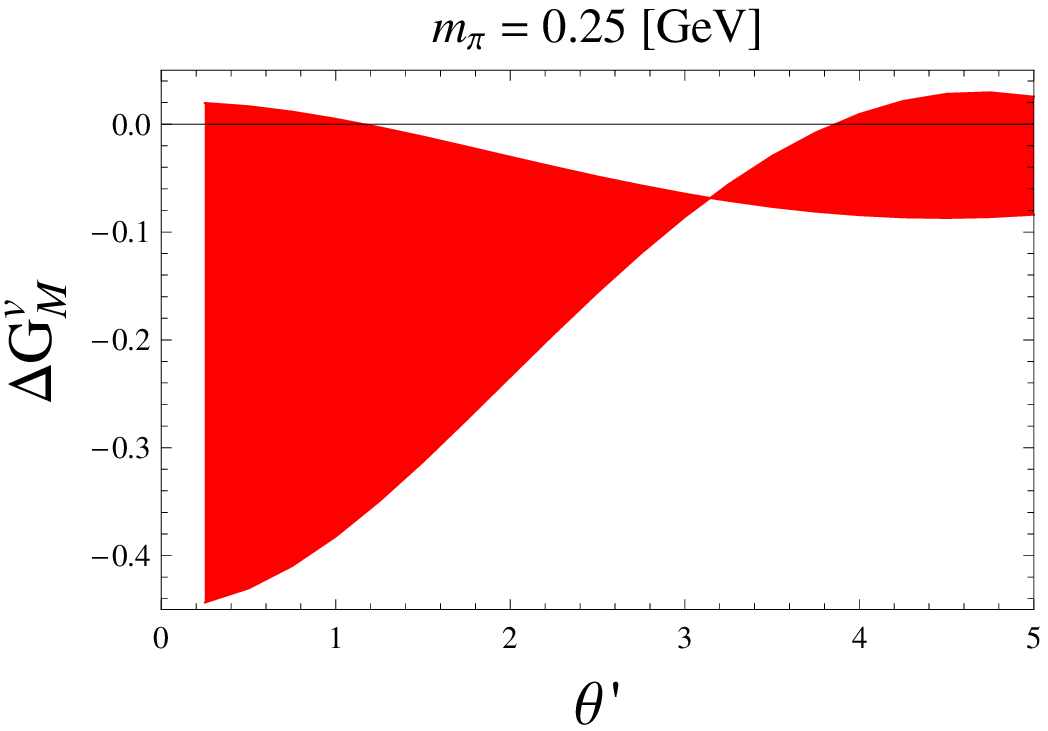,width=3in}
\epsfig{file=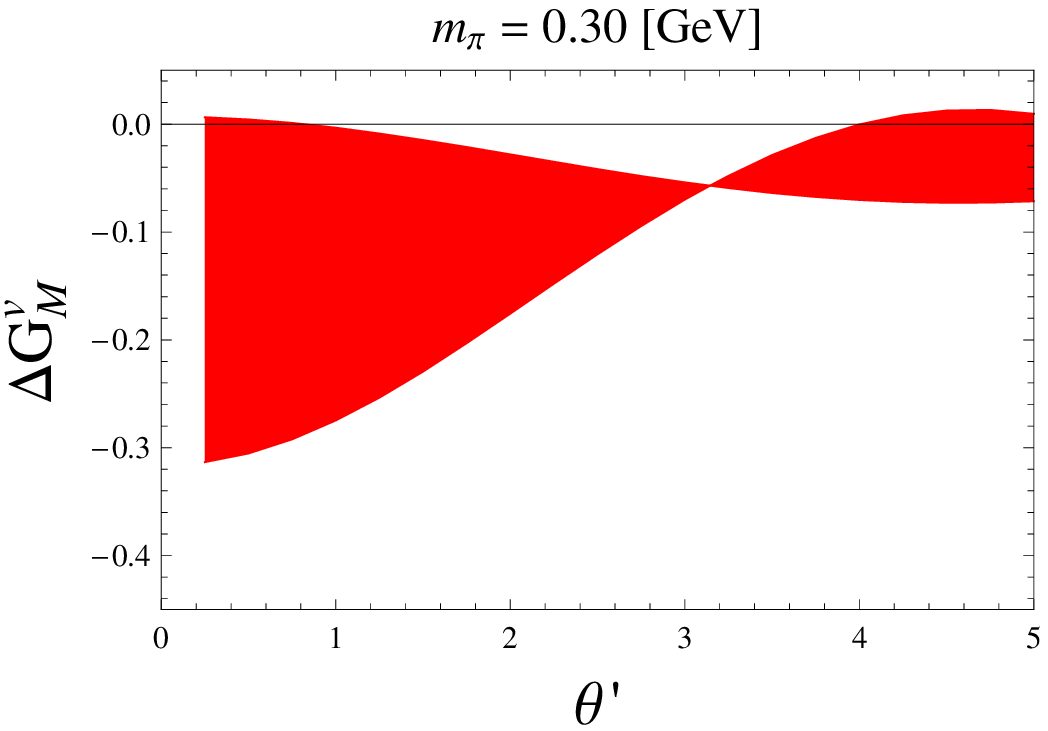,width=3in}
\epsfig{file=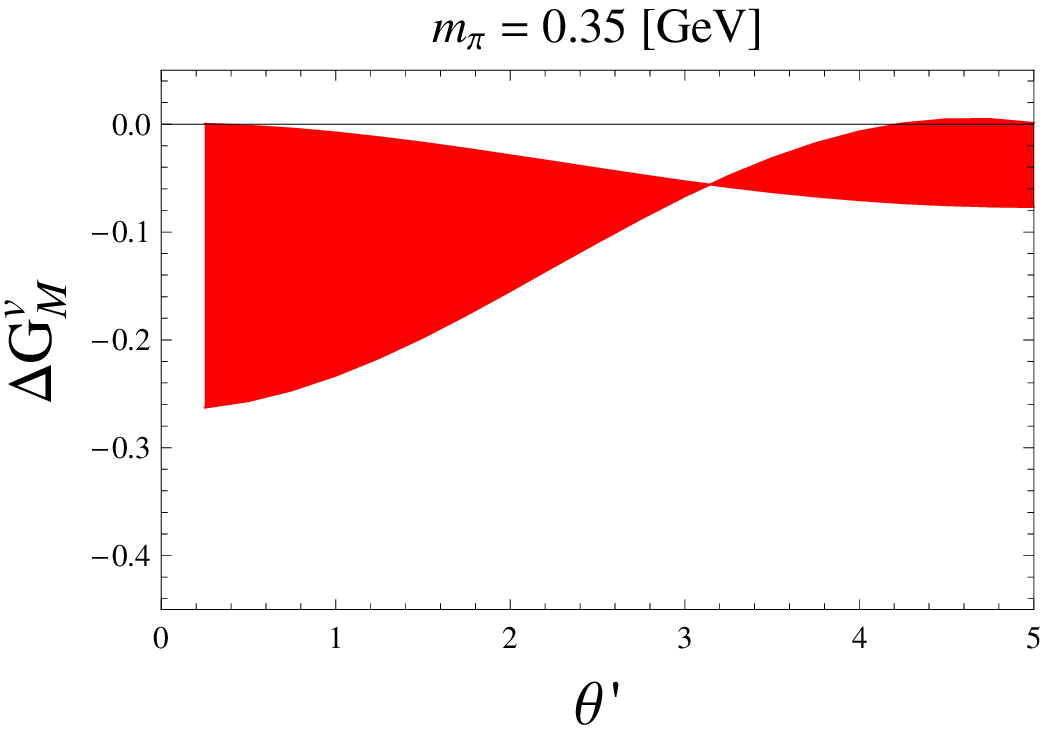,width=3in}
\caption{
Relative change in the isovector magnetic form factor due to twisted boundary conditions in the rest frame.
Plotted versus the twisting angle $\theta'$ is $\D G_M^v(\bm{Q}^2,L)$ given in Eq.~\eqref{eq:GMrel}.
The lattice size $L$ is fixed at $2.75 \, \texttt{fm}$,
and the momentum transfer is
$|\bm{Q}| = \theta' / L =  \theta' \times 0.072 \, \texttt{GeV}$.
The bands arise from uncertainty in the low-energy constant $g_1$, 
which we vary assuming it is of natural size.
}
\label{f:dGM}
\end{center}
\end{figure}

\begin{figure}
\begin{center}
\epsfig{file=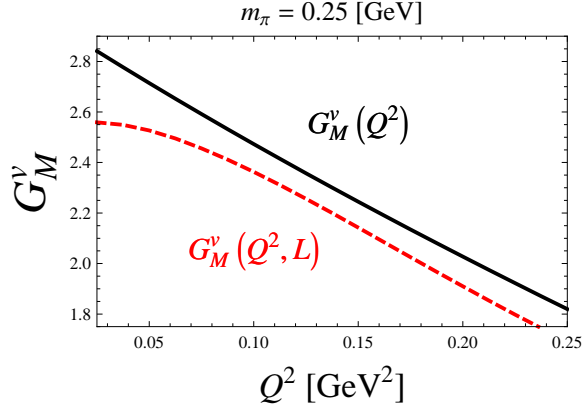,width=3in}
\caption{
Comparison of finite volume and infinite volume isovector magnetic form factors. 
Plotted as functions of $\bm{Q}^2$ are the infinite volume form factor
$G_M^v(\bm{Q}^2)$, 
and finite volume form factor
$G_M^v(\bm{Q}^2, L)$. 
The lattice size is $2.75 \, \texttt{fm}$, and the value of
the unknown axial coupling has been fixed to 
$g_1 = -0.5$. 
}
\label{f:GM}
\end{center}
\end{figure}

Considering the spatial components of the current, 
we can determine the magnetic form factor. 
Additionally there are volume corrections to the 
spatial current, and the net effect has the form 
\begin{equation}
G_M^v(\bm{Q}^2, L) = G^v_M(\bm{Q}^2) + \d_L [ G_M^v(\bm{Q}^2) ]
,\end{equation}
where $G_M^v(\bm{Q}^2)$ is the infinite volume form factor
given by Eq.~\eqref{eq:GMv} and $\d_L [G_M^v(\bm{Q}^2)]$ 
is the finite volume correction, which follows from  Eq.~\eqref{eq:GMvRest}.
Choosing for simplicity $\bm{B}' = B' \hat{\bm{y}}$, and utilizing the
$\hat{\bm{z}}$ component of the current between 
an initial state spin-up and final state spin-down,
we have
\begin{eqnarray}
\d_L [ G_M^v(\bm{Q}^2) ]
&=&
\frac{- 2 M_N}{B' f^2} 
( g_A^2 +  g_A g_1 )
\cK^2(m_\pi, B' \bm{\hat{y}}, 0)
\notag \\
&& + 
\frac{3 M_N}{f^2} 
\int_0^1 dx 
\Big[
g_A^2
\cL^{33} (m_\pi P_\pi, \bm{0}, B' \bm{\hat{y}},  x B' \bm{\hat{y}}, 0) 
\notag \\
&& \phantom{spacespace}
+
\frac{2}{9} g_{\D N}^2
\cL^{33} (m_\pi P_\pi, \bm{0}, B' \bm{\hat{y}}, x B' \bm{\hat{y}}, \D)
\Big].
\end{eqnarray}
Notice this volume correction depends on the unphysical coupling 
$g_1$ which arises as a consequence of having enlarged the valence flavor group. 
The infinite volume isovector magnetic form factor depends upon 
the parameter $\mu_I$ which we can estimate using the known 
values of the proton and neutron magnetic moments. 
We find $\mu_I = 6.77$. 
In Figure~\ref{f:dGM}, we plot the relative change in the 
isovector magnetic form factor due to volume effects
$\D G^v_M$ defined by
\begin{equation} \label{eq:GMrel}
\D G^v_M(\bm{Q}^2, L) = \frac{G_M^v(\bm{Q}^2, L) - G^v_M(\bm{Q}^2)}{G^v_M(\bm{Q}^2)}
.\end{equation}
Again we keep the box size fixed at $2.75 \, \texttt{fm}$, 
and plot versus the twisting angle $\theta'$.
Because the effect is non-negligible, 
we choose a few values of the pion mass. 
The result, moreover, is sensitive to the value of $g_1$
which has been varied assuming natural size, $- 2 \leq g_1 \leq 2$. 
In Figure~\ref{f:GM}, we compare
the extracted form factor at finite volume 
$G_M^v (\bm{Q}^2, L)$ 
with the infinite volume form factor
$G_M^v(\bm{Q}^2)$
as a function of $\bm{Q}^2 = \theta' {}^2 / L^2$. 
In this figure, 
we keep the lattice size at $2.75 \, \texttt{fm}$,
and fix the pion mass to be $0.25 \, \texttt{GeV}$. 
Furthermore, we choose the value of $g_1$
favored by comparing with
$SU(3)$ 
chiral perturbation theory, 
namely 
$g_1 =  2 ( F - D) \approx -0.5$.%
\footnote{  
One could calculate $g_1$ directly by 
determining the axial couplings of hyperons
in the $SU(3)$ limit (and in the chiral regime). 
The first lattice calculation of hyperon axial
charges has been recently performed~\cite{Lin:2007ap},
but naturally with a focus on $SU(3)$ breaking.
}

Lastly we comment on the size of volume corrections in the Breit frame. 
Comparing the expressions for the isovector electric form factor in 
the rest frame Eq.~\eqref{eq:GEvRest}, and the Breit frame Eq.~\eqref{eq:GEvBreit}, 
we see that all factors depending on $\bm{B}$ but not $\bm{Q}$ 
are doubled in the Breit frame. 
This is due to the symmetry under the exchange of the initial 
and final state twists: the finite volume functions are even. 
Because there is some cancellation among the contributions to the 
finite volume electric form factor in the rest frame, we can anticipate
that the finite volume corrections in the Breit frame will generally be of the same size. 
For the magnetic form factor, comparing Eq.~\eqref{eq:GMvRest} and
Eq.~\eqref{eq:GMvBreit} shows similarly that the effect from the $\cK^j$
terms doubles. While this function is odd with respect to argument, terms 
from the initial and final states add coherently because there is a relative
sign from the spin algebra. 
Because empirically we observe the dominant volume correction arises from 
the  $\cK^j$ term, the volume effect for the magnetic form factor will 
roughly double in magnitude in the Breit frame. 
Given that the coefficient of this term depends upon the unphysical 
and unknown parameter $g_1$, the Breit frame does not 
offer an advantage over the rest frame.  
A lattice calculation of $g_1$ is necessary to control the systematic
uncertainty from volume effects in this approach. 
This is not the case for isospin twisted boundary conditions~\cite{Tiburzi:2006px}, 
see Appendix~\ref{A:oldresult}.

\section{Summary} \label{summy}

In this work, we compute finite volume modifications induced by partially twisted boundary conditions.
We utilize heavy baryon chiral perturbation theory in finite volume.
Baryons are embedded into representations of  $SU(7|5)$, where the extra flavors are fictitious,
and differ only in their boundary conditions.
The nucleon mass splittings are determined, and demonstrated to be negligible on current-sized lattices.
The main focus of our work is the derivation of finite volume corrections to the vector current 
matrix elements of the nucleon. 
Continuous momentum is inserted on the active valence quark lines using flavor changing currents in the enlarged flavor group.
Disconnected operator insertions cannot be accessed at continuous momentum using this technique, 
and our calculation is therefore restricted to connected current insertions.
Isospin breaking and cubic symmetry breaking lead to various structures not encountered in infinite volume. 
We give complete expressions for finite volume current matrix elements using  general kinematics.
To estimate the size of these corrections, we choose rest frame kinematics,
and consider both the spatial and temporal components of the current.
Generally the volume corrections lead to oscillatory behavior about the infinite volume answer. 
In the region of small twist angles, the volume effects can become rather pronounced
due to terms that break cubic symmetry. 
To extract the isovector magnetic moment and electromagnetic radii from lattice data 
at zero Fourier momentum, 
careful determination of volume effects will be required.
This is complicated by the dependence on an unphysical and unknown axial coupling $g_1$.  
As shown in Appendix~\ref{A:oldresult}, a different implementation of twisted boundary 
conditions can eliminate this dependence. In this implementation, there are no fictitious flavors, 
rather the isospin transition is simulated directly. 
Compared to the meson sector, the baryon sector appears more susceptible to volume corrections due to partial twisting.
The Breit frame kinematics do not simplify or reduce the volume corrections. 
Partial twisting provides a novel way to probe any isovector nucleon matrix element
at continuous values of momentum transfer. 
The formalism developed here allows one to compute
finite volume modifications to these observables, 
and thereby control the extraction of moments, radii, \emph{etc}, from lattice QCD data.

\begin{acknowledgments}
This work is supported in part by the 
Schweizerischer Nationalfonds,
and by the
U.S.~Dept.~of Energy,
Grant No.~DE-FG02-93ER-40762.
B.C.T. acknowledges the Institute for Nuclear Theory at the University of Washington 
for its hospitality during the initial stages of this work. 
\end{acknowledgments}

\appendix

\section{Finite Volume Current Matrix Elements}
\label{A:result}

In this Appendix, we list the finite volume corrections to current matrix elements. 
For ease of presentation, we remove the Pauli spinors. 
Here we work in a general frame where the initial-state nucleon has
momentum $\bm{B} = \bm{\theta}/L$, and the final-state nucleon has
momentum $\bm{B}' + \bm{q}$, where $\bm{q}$ is a Fourier momentum 
mode of the lattice and $\bm{B}' = \bm{\theta'}/L$. 
Here we list only the proton matrix elements as a function of $q_u$ and $q_d$. 
One can use charge symmetry, $q_u \leftrightarrow q_d$, to deduce 
the neutron matrix elements. 
As explained in the main text, 
each result includes only connected part of the matrix elements.

The finite volume modification to the time-component of the current matrix element
in Eq.~\eqref{eq:recipe} reads
{ 
\scriptsize
\begin{eqnarray}
\d J_4
&=&  
\frac{1}{f^2}
\int_0^1 dx 
(2 q_u + q_d )
\left[ 
\cI_{1/2} (m_{ju} P_{ju}, x \bm{Q} + \bm{B})
- 
\frac{1}{2} \cI_{1/2} (m_{ju},\bm{B})
- 
\frac{1}{2} \cI_{1/2} (m_{ju},\bm{B}')
\right] 
\notag \\
&& - 
\frac{3}{2  f^2}
\Bigg\{ 
\left(  q_d +  \frac{1}{2} q_u  \right)
\Bigg[
g_{\pi N_{\bm{3}} N_{\bm{3}}}^{2} 
\ol \cJ(m_\pi, \bm{0},0) 
+ 
g_{ju N_{\bm{3}} N_{\bm{3}}}^{2} 
\ol \cJ(m_{ju}, \bm{0},0) 
+
\frac{1}{2} 
g_{\pi N_{\bm{3}} N_{\bm{3}}}^{\prime \, 2} 
\Big( \ol \cJ(m_\pi, \bm{B}, 0)  + \ol \cJ(m_\pi, \bm{B}', 0) \Big)
\notag\\
&& \phantom{spaspaspa} 
+
\frac{1}{2} 
g_{ju N_{\bm{3}} N_{\bm{3}}}^{\prime \, 2} 
\Big( \ol  \cJ(m_{ju}, \bm{B}, 0)  + \ol \cJ(m_{ju}, \bm{B}', 0) \Big)
\Bigg]  
\notag \\
&& \phantom{spa}
+ 
\frac{3}{2} q_u 
\Bigg[
g_{\pi N_{\bm{1}} N_{\bm{1}}}^{2} 
\ol  \cJ(m_\pi, \bm{0},0) 
+
g_{ju N_{\bm{1}} N_{\bm{1}}}^{2} 
\ol \cJ(m_{ju}, \bm{0},0) 
+ 
\frac{1}{2}  
g_{\pi N_{\bm{1}} N_{\bm{1}}}^{\prime \, 2} 
\Big( 
\ol \cJ(m_\pi, \bm{B}, 0)  + \ol \cJ(m_\pi, \bm{B}', 0) \Big)
\notag \\
&& \phantom{spaspaspa}
+ 
\frac{1}{2}  
g_{ju N_{\bm{1}} N_{\bm{1}}}^{\prime \, 2} 
\Big( \ol  \cJ(m_{ju}, \bm{B}, 0)  + \ol \cJ(m_{ju}, \bm{B}', 0) \Big)
\Bigg]
\notag \\
&& \phantom{spaspa}
+
\left( q_d + \frac{1}{2} q_u \right) 
\frac{g_{\D N}^2}{9} 
\Bigg[
\ol \cJ(m_\pi, \bm{0},\D) 
+ 
2 \ol \cJ(m_{ju}, \bm{0},\D) 
+
\frac{5}{2} 
\Big( 
\ol \cJ(m_\pi, \bm{B},\D) 
+
\ol \cJ(m_\pi, \bm{B}',\D) 
\Big)
\notag \\
&& \phantom{spaspaspa}
+
2
\Big( 
\ol \cJ(m_{ju}, \bm{B},\D) 
+
\ol \cJ(m_{ju}, \bm{B}',\D) 
\Big)
\Bigg]
\notag \\
&& \phantom{spa}
+
\left(
\frac{3}{2}  q_u  
\right)
\frac{g_{\D N}^2}{3}
\Bigg[
\ol \cJ(m_\pi, \bm{0},\D) 
+ 
2 \ol \cJ(m_{ju}, \bm{0},\D) 
+ 
\frac{1}{2}
\Big( 
\ol \cJ(m_\pi, \bm{B},\D) 
+
\ol \cJ(m_\pi, \bm{B}',\D) 
\Big)
\Bigg]
\Bigg\}
\notag \\
&& +
\frac{3}{2 f^2}
\Bigg\{
\left(  q_d + \frac{1}{2} q_u \right) 
\left[
g_{\pi N_{\bm{3}} N_{\bm{3}}}^{2} 
\ol \cJ(m_{\pi}, \bm{0},0) 
+
\frac{1}{3}(g_A^2 - g_A g_1 - g_1^2 /2)
\Big( 
\ol \cJ(m_\pi, \bm{B},0) 
+
\ol \cJ(m_\pi, \bm{B}',0) 
\Big)
+
g_{ju N_{\bm{3}} N_{\bm{3}}}^{2} 
\ol \cJ(m_{ju}, \bm{0},0) 
\right]
\notag \\
&& \phantom{spaspaspa}
+
\frac{3}{2} q_u 
\left[
g_{\pi N_{\bm{1}} N_{\bm{1}}}^{2} 
\ol \cJ (m_\pi, \bm{0}, 0)
+
\frac{1}{9} (g_A^2 + 5 g_A g_1 - g_1^2 / 2)
\Big( 
\ol \cJ(m_\pi, \bm{B},0) 
+
\ol \cJ(m_\pi, \bm{B}',0) 
\Big)
+
g_{ju N_{\bm{1}} N_{\bm{1}}}^{2} 
\ol \cJ(m_{ju}, \bm{0},0) 
\right]
\notag \\
&& \phantom{spaspa}
+
g_{\D N}^2
\Bigg[
\left(  q_d + \frac{1}{2} q_u \right) 
\left( 
\frac{1}{9} 
\ol \cJ(m_{\pi}, \bm{0},\D) 
+ 
\frac{1}{9}
\Big( 
\ol \cJ(m_\pi, \bm{B},\D) 
+
\ol \cJ(m_\pi, \bm{B}',\D) 
\Big)
+
\frac{2}{9}
\ol \cJ(m_{ju}, \bm{0},\D) 
\right)
\notag \\
&& \phantom{spaspaspa}
+ 
\frac{3}{2} q_u 
\left( 
\frac{1}{3}
\ol \cJ(m_{\pi}, \bm{0},\D) 
+ 
\frac{1}{3}
\Big( 
\ol \cJ(m_\pi, \bm{B},\D) 
+
\ol \cJ(m_\pi, \bm{B}',\D) 
\Big)
+
\frac{2}{3}
\ol \cJ(m_{ju}, \bm{0},\D) 
\right)
\Bigg]
\Bigg\}
\notag \\
&& -
\frac{3}{ f^2} 
[ \bm{S} \cdot \bm{Q}, S_j ]
\int_0^1 dx 
\left[ 
\b_\pi 
\cJ^j ( m_\pi P_\pi, \bm{B}, x \bm{Q} + \bm{B}, 0 )
+ 
\b_{ju} 
\cJ^j (m_{ju} P_{ju}, \bm{B}, x \bm{Q} + \bm{B}, 0)
\right]
\notag \\
&& - 
\frac{3 g_{\D N}^2 }{f^2}
[ \bm{S} \cdot \bm{Q}, S_j ]
\int_0^1 dx 
\left[ 
\b'_\pi 
\cJ^j ( m_\pi P_\pi, \bm{B}, x \bm{Q} + \bm{B}, \D )
+ 
\b'_{ju} 
\cJ^j (m_{ju} P_{ju}, \bm{B}, x \bm{Q} + \bm{B}, \D)
\right]
\notag \\
&& -
\frac{3}{2 f^2}
\int_0^1 dx 
\left[ 
\b_\pi
\cJ(m_{\pi} P_{\pi}, \bm{B}, \bm{Q} + \bm{B}, x \bm{Q} + \bm{B}, 0)
+ 
\b_{ju} 
\cJ(m_{ju} P_{ju}, \bm{B}, \bm{Q} + \bm{B}, x \bm{Q} + \bm{B}, 0)
\right] 
\notag \\
&& + 
\frac{3 g_{\D N}^2 }{f^2}
\int_0^1 dx 
\left[ 
\b_\pi' 
\cJ(m_{\pi} P_{\pi}, \bm{B}, \bm{Q} + \bm{B}, x \bm{Q}  + \bm{B}, \D)
+ 
\b_{ju}' 
\cJ(m_{ju} P_{ju}, \bm{B}, \bm{Q} + \bm{B}, x \bm{Q}  +  \bm{B}, \D)
\right].
\end{eqnarray}
}%
The effective axial couplings have been given above in Eqs.~\eqref{eq:N3axial} and
\eqref{eq:N1axial}, while the loop coefficients appear in Eqs.~\eqref{eq:beta} and \eqref{eq:betaprime}. 
The spatial components of the current matrix element in Eq.~\eqref{eq:recipe}
receive the finite volume modification
{\scriptsize
\begin{eqnarray}
\d J_{i}
&=& 
- \frac{1}{ 2 f^2}
\Bigg\{
\left( q_d + \frac{1}{2} q_u \right)
\Big[
g^{\prime \, 2}_{\pi N_{\bm{3}} N_{\bm{3}}}
\Big(
\cK^i(m_{\pi}, \bm{B}, 0) + \cK^i(m_{\pi}, \bm{B}', 0)
\Big)
+
g^{\prime \, 2}_{ju N_{\bm{3}} N_{\bm{3}}}
\Big(
\cK^i(m_{ju}, \bm{B}, 0) + \cK^i(m_{ju}, \bm{B}', 0)
\Big)
\Big]
\notag \\
&& \phantom{space}+ 
\frac{3}{2} q_u
\Big[
g^{\prime \, 2}_{\pi N_{\bm{1}} N_{\bm{1}}}
\Big(
\cK^i(m_{\pi}, \bm{B}, 0) + \cK^i(m_{\pi}, \bm{B}', 0)
\Big)
+
g^{\prime \, 2}_{ju N_{\bm{1}} N_{\bm{1}}}
\Big(
\cK^i(m_{ju}, \bm{B}, 0) + \cK^i(m_{ju}, \bm{B}', 0)
\Big)
\Big]
\Bigg\}
\notag \\
&& -
\frac{[S_i, S_j]}{f^2} 
\Bigg\{
\left( q_d + \frac{1}{2} q_u \right)
\Big[
g^{\prime \, 2}_{\pi N_{\bm{3}} N_{\bm{3}}}
\Big(
\cK^j(m_{\pi}, \bm{B}, 0) - \cK^j(m_{\pi}, \bm{B}', 0)
\Big)
+
g^{\prime \, 2}_{ju N_{\bm{3}} N_{\bm{3}}}
\Big(
\cK^j(m_{ju}, \bm{B}, 0) - \cK^j(m_{ju}, \bm{B}', 0)
\Big)
\Big]
\notag \\
&& \phantom{space}+ 
\frac{3}{2} q_u
\Big[
g^{\prime \, 2}_{\pi N_{\bm{1}} N_{\bm{1}}}
\Big(
\cK^j(m_{\pi}, \bm{B}, 0) - \cK^j(m_{\pi}, \bm{B}', 0)
\Big)
+
g^{\prime \, 2}_{ju N_{\bm{1}} N_{\bm{1}}}
\Big(
\cK^j(m_{ju}, \bm{B}, 0) - \cK^j(m_{ju}, \bm{B}', 0)
\Big)
\Big]
\Bigg\}
\notag \\
&&
-\frac{g_{\D N}^2}{ 3 f^2}
\Bigg\{
\left( q_d + \frac{1}{2} q_u \right)
\Big[
\frac{5}{6}
\Big(
\cK^i(m_{\pi}, \bm{B}, \D) + \cK^i(m_{\pi}, \bm{B}', \D)
\Big)
+
\frac{2}{3}
\Big(
\cK^i(m_{ju}, \bm{B}, \D) + \cK^i(m_{ju}, \bm{B}', \D)
\Big)
\Big]
\notag \\
&& \phantom{space}+ 
\left( \frac{3}{2} q_u \right)
\frac{1}{2}
\Big(
\cK^i(m_{\pi}, \bm{B}, \D) + \cK^i(m_{\pi}, \bm{B}', \D)
\Big)
\Bigg\}
\notag \\
&& +
g_{\D N}^2
\frac{[S_i, S_j]}{3 f^2} 
\Bigg\{
\left( q_d + \frac{1}{2} q_u \right)
\Big[
\frac{5}{6}
\Big(
\cK^j(m_{\pi}, \bm{B}, \D) - \cK^j(m_{\pi}, \bm{B}', \D)
\Big)
+
\frac{2}{3}
\Big(
\cK^j(m_{ju}, \bm{B}, \D) - \cK^j(m_{ju}, \bm{B}', \D)
\Big)
\Big]
\notag \\
&& \phantom{space}+ 
\left( \frac{3}{2} q_u \right)
\frac{1}{2}
\Big(
\cK^j(m_{\pi}, \bm{B}, \D) - \cK^j(m_{\pi}, \bm{B}', \D)
\Big)
\Bigg\}
\notag \\
&& -
\frac{3}{4f^2}
\int_0^1 dx 
\left[
\b_\pi 
\cL^i (m_\pi P_\pi, \bm{B}, \bm{Q} + \bm{B}, x \bm{Q} + \bm{B}, 0) 
+
\b_{ju} 
\cL^i (m_{ju} P_{ju}, \bm{B}, \bm{Q} + \bm{B}, x \bm{Q} + \bm{B}, 0) 
\right]
\notag \\
&& +
\frac{3 g_{\D N}^2}{2f^2}
\int_0^1 dx 
\left[
\b'_\pi 
\cL^i (m_\pi P_\pi, \bm{B}, \bm{Q} + \bm{B}, x \bm{Q} + \bm{B}, \D) 
+
\b'_{ju} 
\cL^i (m_{ju} P_{ju}, \bm{B}, \bm{Q} + \bm{B}, x \bm{Q} + \bm{B}, \D) 
\right]
\notag \\
&& - 
\frac{3}{2 f^2} 
[\bm{Q} \cdot \bm{S}, S_j]
\int_0^1 dx 
\left[
\beta_\pi 
\cL^{ji} (m_\pi P_\pi, \bm{B}, \bm{Q} +  \bm{B}, x \bm{Q} + \bm{B}, 0)
+ 
\beta_{ju}
\cL^{ji} (m_{ju} P_{ju}, \bm{B}, \bm{Q} +  \bm{B}, x \bm{Q} + \bm{B}, 0)
\right]
\notag \\
&& - 
\frac{3 g_{\D N}^2}{2 f^2} 
[\bm{Q} \cdot \bm{S}, S_j]
\int_0^1 dx 
\left[
\beta'_\pi 
\cL^{ji} (m_\pi P_\pi, \bm{B}, \bm{Q} +  \bm{B}, x \bm{Q} + \bm{B}, \D)
+ 
\beta'_{ju}
\cL^{ji} (m_{ju} P_{ju}, \bm{B}, \bm{Q} +  \bm{B}, x \bm{Q} + \bm{B}, \D)
\right]
.\end{eqnarray}
}%

Appearing in the above expressions for finite volume modifications 
are functions depending on the difference of 
finite volume mode sums and infinite volume momentum integrals. 
The various definitions are as follows:
{\footnotesize
\begin{eqnarray}
\cI_{1/2}(m, \bm{A}) 
&=& 
\frac{1}{L^3}
\sum_{\bm{n}}\frac{1}{[(\bm{k} + \bm{A})^2 + m^2]^{1/2}}
- 
\int \frac{d \bm{k}}{(2 \pi)^3} \frac{1}{[\bm{k}^2 + M^2]^{1/2}}
,\end{eqnarray}
\begin{equation}
\cJ (m, \bm{A}, \bm{B}, \bm{C}, \D)
= 
\int_0^\infty  d\l
\, \l \left[ 
\frac{1}{L^3}
\sum_{\bm{n}}
\frac{( \bm{k} + \bm{A}) \cdot ( \bm{k} + \bm{B})}
{[(\bm{k} + \bm{C})^2 + \b_\D^2]^{5/2}} 
- 
\int \frac{d \bm{k}}{(2 \pi)^3} \frac{(\bm{k} + \bm{A}) \cdot (\bm{k} + \bm{B})}
{[(\bm{k}+\bm{C})^2 + \b_\D^2]^{5/2}}\right]
,\end{equation}
\begin{equation}
\cJ {}^j (m, \bm{A}, \bm{B}, \D)
= 
\int_0^\infty  d\l \, \l 
\left[ 
\frac{1}{L^3}
\sum_{\bm{n}}
\frac{(\bm{k} + \bm{A})^j}
{[(\bm{k} + \bm{B})^2 + \b_\D^2]^{5/2}} 
- 
\int \frac{d \bm{k}}{(2 \pi)^3} 
\frac{(\bm{k} + \bm{A})^j }{[(\bm{k}+ \bm{B})^2 + \b_\D^2]^{5/2}}\right]
,\end{equation}
\begin{equation}
\cK^j (m, \bm{B}, \D) 
=
\int_0^\infty  d\l
\frac{1}{L^3}
\sum_{\bm{n}}
\frac{(\bm{k} + \bm{B})^j}{[(\bm{k} + \bm{B})^2 + \b_\D^2]^{3/2}}
,\end{equation}
\begin{eqnarray}
\cL^{j} (m, \bm{A}, \bm{B}, \bm{C}, \D)
&=& 
\int_0^\infty  d\l
\Bigg[ \frac{1}{L^3}
\sum_{\bm{n}}
\frac{(\bm{k} + \bm{A}) \cdot ( \bm{k} + \bm{B})   ( 2 \bm{k} + \bm{A}+ \bm{B})^j}
{[(\bm{k} + \bm{C})^2 + \b_\D^2]^{5/2}} 
\notag \\
&& \phantom{spacespace}
- 
\int \frac{d \bm{k}}{(2 \pi)^3} \frac{(\bm{k}+ \bm{A}) \cdot ( \bm{k} + \bm{B})  (2 \bm{k} + \bm{A} + \bm{B})^j}{[(\bm{k}+ \bm{C})^2 + \b_\D^2]^{5/2}}
\Bigg]
,\end{eqnarray}
and
\begin{equation}
\cL^{ij} (m, \bm{A}, \bm{B}, \bm{C}, \D)
= 
\int_0^\infty  d\l
\left[ \frac{1}{L^3}
\sum_{\bm{n}}
\frac{(\bm{k} + \bm{A})^i ( 2 \bm{k} + \bm{A} + \bm{B})^j}
{[(\bm{k} + \bm{C})^2 + \b_\D^2]^{5/2}} 
- 
\int \frac{d \bm{k}}{(2 \pi)^3} \frac{(\bm{k}+ \bm{A})^i (2 \bm{k} +  \bm{A} + \bm{B})^j}{[(\bm{k}+ \bm{C})^2 + \b_\D^2]^{5/2}}\right]
.\end{equation}
}%
We also use the short hand $\ol \cJ  (m, \bm{A}, \D) \equiv \cJ (m, \bm{A}, \bm{A}, \bm{A}, \D)$.
We show how to evaluate these functions numerically in Appendix~\ref{s:FVS}.

\section{Finite Volume Isovector Current Matrix Elements from $SU(4|2)$}
\label{A:oldresult}

In this Appendix, 
we detail the finite volume corrections to current matrix elements
using the alternate implementation of partially twisted boundary conditions proposed in~\cite{Tiburzi:2005hg,Tiburzi:2006px}. 
This method does not rely on fictitious flavors of valence quarks. 
Instead, one directly confronts the isospin changing operators whose matrix elements give rise to isovector form factors. 
From the outset, one is aware that disconnected contributions cannot be accessed. 
The flavor structure, moreover, only requires a simple modification of existing partially quenched theories to include twisted boundary conditions. 
Results for the finite volume isovector magnetic form factor under simplifying kinematics were given in~\cite{Tiburzi:2006px}; 
however, 
as complete expressions for finite volume current matrix elements were not presented, we give the complete expressions here.

Let us briefly summarize the setup used in~\cite{Tiburzi:2006px}.
We restrict our attention to an $SU(4|2)$ theory with quarks contained 
in a field 
$Q$, 
which is given by
$Q = ( u, d, j, l, \tilde{u}, \tilde{d} )^T$. 
Each quark is periodic but coupled to a uniform Abelian gauge potential $B_\mu$
of the form
$B_\mu = \diag ( B_\mu^u, B_\mu^d, 0, 0, B_\mu^u, B_\mu^d )$, 
with 
$B_\mu^u = (\bm{\theta}^u / L, 0)$ 
and
$B_\mu^d = (\bm{\theta}^d / L, 0)$.  
In this formulation, 
momentum is injected by isospin changing operators provided 
$B_\mu^d \neq B_\mu^d$.
Keeping these twists different introduces isospin breaking via finite volume effects. 
The partially quenched isospin splittings for the pion were numerically demonstrated
to be quite small on current lattices~\cite{Jiang:2006gna}. 
To calculate the nucleon isospin splitting, 
we evaluate the sunset diagrams shown in Figure~\ref{f:Nmass}
in the partially twisted $SU(4|2)$ theory. 
The nucleon isospin splitting is given by
\begin{eqnarray}
M_n - M_p 
&=&
- 
\frac{1}{2 f^2} 
\Bigg\{ 
\frac{1}{6} 
g_{ju NN}^2
\big[ \cK(m_{ju}, \bm{B}_d, 0) - \cK(m_{ju}, \bm{B}_u, 0) \big]
\notag \\
&& 
\phantom{space} - \frac{2}{9} g_{\D N}^2 \big[ \cK(m_{ju}, \bm{B}_d, \D) - \cK(m_{ju}, \bm{B}_u, \D)  \big] 
\Bigg\}
.\end{eqnarray}
The effective axial coupling $g_{ju NN}^2$ has been given in Eq.~\eqref{eq:PQcouplings}. 
When the twists are isospin symmetric, the nucleon mass splitting accordingly vanishes. 
The maximal isospin splitting is occurs when $\bm{B} = \pi (1,1,1)$ for one flavor, and 
$\bm{B} = \bm{0}$ for the other. On current lattices this maximal splitting is at the percent level
and can practically be ignored.

For the operator $J^+_\mu = \ol u \gamma_\mu d$, continuous 
three-momentum of the form $\bm{B}_{\pi} = \bm{B}^u - \bm{B}^d$
is induced in flavor changing matrix elements. 
Thus we consider the isovector-vector current matrix elements between
nucleons
\begin{eqnarray} \label{eq:isovector}
\langle p( \bm{q} ) | J^+_\mu | n(\bm{0}) \rangle
\overset{L \to \infty}{\longrightarrow}
\langle p( \bm{P'} ) | J^+_\mu | n(\bm{P}) \rangle
=
\langle p(\bm{P'}) | J^{\text{em}}_\mu | p(\bm{P}) \rangle 
- 
\langle n(\bm{P'}) | J^{\text{em}}_\mu | n(\bm{P}) \rangle
.\end{eqnarray}
On the left, we have denoted only the Fourier momentum. 
On the right, 
the momentum of the initial-state nucleon due to twisting is 
$\bm{P} = \bm{B}^u + 2 \bm{B}^d$,
while the final-state nucleon has momentum
$\bm{P'} = \bm{q} + 2 \bm{B}^u + \bm{B}^d$. 
The momentum transfer we denote by $\bm{Q}$
and is given here by $\bm{Q} = \bm{q} + \bm{B}_\pi$.
The equality between the isovector-vector current and
differences of the electromagnetic current matrix elements 
follows from the 
$SU(2)_{\text{valence}}$ 
symmetry subgroup of the full 
$SU(4|2)$ 
group. 
At finite volume, this symmetry is broken and one must 
address the volume corrections to the matrix element on the right-hand side. 
To determine these corrections, 
we evaluate the one-loop diagrams for the isospin transition matrix element
using the partially twisted 
$SU(4|2)$ 
theory.
The relevant diagrams are shown in Figures~\ref{f:Nvec} and \ref{f:moreNvec}.

The finite volume modification to the time-component of the 
isovector current matrix element in Eq.~\eqref{eq:isovector} reads
{\footnotesize
\begin{eqnarray}
\d J^{+}_4
&=&  
\frac{1}{f^2}
\int_0^1 dx 
\left[ 
\cI_{1/2} (m_{ju} P_{ju}, x \bm{Q} + \bm{B}_d)
- 
\frac{1}{2} \cI_{1/2} (m_{ju},\bm{B}_u)
- 
\frac{1}{2} \cI_{1/2} (m_{ju},\bm{B}_d)
\right] 
\notag \\
&& - 
\frac{1}{8  f^2}
\Bigg\{ g_{\pi NN}^2 \Big[\ol \cJ(m_\pi, \bm{0},0) + \ol \cJ(m_\pi, \bm{B}_\pi, 0) \Big]  
+ g_{ju NN}^2 \Big[\ol \cJ(m_{ju}, \bm{B}_u,0) + \ol \cJ(m_{ju}, \bm{B}_d, 0)\Big]  
\Bigg\}
\notag \\
&& + 
\frac{g_{\D N}^2}{6 f^2} 
\Big[ 
3 \ol \cJ ( m_\pi, \bm{0}, \D) 
+ 
3 \ol \cJ (m_\pi, \bm{B}_\pi, \D) 
+ 
\ol \cJ (m_{ju}, \bm{B}_u, \D) 
+
\ol \cJ(m_{ju}, \bm{B}_d, \D)
\Big] 
\notag \\
&& +
\frac{ 1}{2  f^2} [\bm{Q} \cdot \bm{S}, S_j]
\int_0^1 dx 
\left[
g_{\pi NN}^2 
\cJ^j (m_{\pi} P_{\pi}, \bm{0}, x \bm{Q} , 0) 
+
g_{ju NN}^2 
\cJ^j (m_{ju} P_{ju}, \bm{B}_d, x \bm{Q} + \bm{B}_d , 0)
\right] 
\notag \\
&&
+ 
\frac{g_{\D N}^2}{f^2}
[\bm{Q} \cdot \bm{S}, S_j]
\int_0^1 dx 
\left[
\cJ^j (m_{\pi} P_{\pi}, \bm{0}, x \bm{Q} , \D) 
+
\frac{1}{3}
\cJ^j (m_{ju} P_{ju}, \bm{B}_d, x \bm{Q} + \bm{B}_d, \D)
\right]
\notag \\
&& +
\frac{1}{4 f^2}
\int_0^1 dx 
\left[ 
g_{\pi NN}^2 
\cJ(m_\pi P_\pi, \bm{0}, \bm{Q},  x \bm{Q}, 0)
+ 
g_{ju NN}^2 
\cJ(m_{ju} P_{ju}, \bm{B}_d, \bm{Q}+ \bm{B}_d,  x \bm{Q} + \bm{B}_d, 0)
\right] 
\notag \\
&& -  
\frac{g_{\D N}^2}{f^2}
\int_0^1 dx 
\left[ 
\cJ(m_\pi P_\pi,  \bm{0}, \bm{Q}, x \bm{Q}, \D)
+ 
\frac{1}{3} 
\cJ(m_{ju} P_{ju}, \bm{B}_d, \bm{Q}+ \bm{B}_d,  x \bm{Q} + \bm{B}_d, \D)
\right].
\end{eqnarray}
}%
We have omitted writing the Pauli spinors here and below. 
The effective axial couplings, 
$g_{\pi NN}^2$ and $g_{ju NN}^2$, 
have been given above, Eq.~\eqref{eq:PQcouplings}. 
The spatial components of the current matrix element in Eq.~\eqref{eq:recipe}
receive the finite volume modification
{\footnotesize
\begin{eqnarray}
\d J^{+}_{i}
&=& 
-
\frac{1}{ 2 f^2}
\Bigg\{
\frac{1}{6} g_{ju NN}^2 
\Big[ 
\cK^i(m_{ju}, \bm{B}_u, 0) + \cK^i(m_{ju}, \bm{B}_d, 0)
\Big]
- \frac{2}{9} g_{\D N}^2
\Big[ 
\cK^i(m_{ju}, \bm{B}_u, \D) + \cK^i(m_{ju}, \bm{B}_d, \D)
\Big]
\Bigg\} 
\notag \\
&& +
\frac{[S_i, S_j]}{6 f^2} 
\Bigg\{ 
g_{\pi NN}^2 \cK^j(m_\pi, \bm{B}_\pi, 0)
+
g_{ju NN}^2  
\Big[ 
\cK^j(m_{ju}, \bm{B}_u, 0) - \cK^j(m_{ju}, \bm{B}_d, 0)
\Big]
\Bigg\} 
\notag \\
&& + 
\frac{g_{\D N}^2 \, [S_i, S_j]}{3 f^2}
\Bigg\{ 
\cK^j(m_\pi, \bm{B}_\pi, \D)
+ 
\frac{1}{3} 
\Big[
\cK^j(m_{ju}, \bm{B}_u, \D) - \cK^j(m_{ju}, \bm{B}_d, \D)
\Big]
\Bigg\}
\notag \\
&& +
\frac{1}{8  f^2}
\int_0^1 dx 
\left[ 
g_{\pi NN}^2
\cL^i (m_{\pi} P_{\pi}, \bm{0}, \bm{Q}, x \bm{Q}, 0)
+
g_{ju NN}^2
\cL^i (m_{ju} P_{ju}, \bm{B}_d, \bm{Q} + \bm{B}_d, x \bm{Q} + \bm{B}_d, 0)
\right]
\notag \\
&&
- 
\frac{g_{\D N}^2}{2 f^2}
\int_0^1 dx
\left[
\cL^i (m_{\pi} P_{\pi}, \bm{0}, \bm{Q}, x \bm{Q}, \D)
+
\frac{1}{3} 
\cL^i (m_{ju} P_{ju}, \bm{B}_d, \bm{Q}+ \bm{B}_d, x \bm{Q} + \bm{B}_d, \D)
\right] 
\notag \\
&& + 
\frac{1}{4 f^2}[\bm{Q} \cdot \bm{S}, S_j]
\int_0^1 dx 
\left[
g_{\pi NN}^2 \cL^{ji} (m_\pi P_\pi, \bm{0}, \bm{Q},  x \bm{Q}, 0) 
+ 
g_{ju NN}^2 
\cL^{ji} (m_{ju} P_{ju}, \bm{B}_d, \bm{Q} +  \bm{B}_d, x \bm{Q} + \bm{B}_d, 0)
\right]
\notag \\
&& + 
\frac{g_{\D N}^2 }{2  f^2} [\bm{Q} \cdot \bm{S}, S_j]
\int_0^1 dx 
\left[
\cL^{ji} (m_\pi P_\pi, \bm{0}, \bm{Q}, x \bm{Q}, \D)
+ 
\frac{1}{3} 
\cL^{ji} (m_{ju} P_{ju}, \bm{B}_d, \bm{Q}+  \bm{B}_d, x \bm{Q} + \bm{B}_d, \D)
\right].
\notag \\
\end{eqnarray}
}%

From these expressions, we can simplify things by forming unpolarized (polarized)
matrix elements for the temporal (spatial) part of the current. 
Furthermore, we restrict our attention to the  unitary mass point, where 
$m_{ju}^2 = m_\pi^2$, 
and choose the twist parameters such that
$\bm{\theta}^d = \bm{0}$
and 
$\bm{\theta}^u = \bm{\theta}$, 
which corresponds to rest frame kinematics.%
\footnote{
The choice $\bm{\theta}^d = - \bm{\theta}^u$
does not result in any dramatic simplifications. 
In particular there will be residual $g_1$ dependence in 
this case. 
}
The time component of the current becomes
{\footnotesize
\begin{eqnarray}
\frac{1}{2} 
\sum_{m = \pm}
\langle m | \d J^{+}_4 | m \rangle
&=&  
\frac{1}{f^2}
\int_0^1 dx 
\left[ 
\cI_{1/2} (m_{\pi} P_{\pi}, x \bm{Q})
- 
\frac{1}{2} \cI_{1/2} (m_{\pi},\bm{0})
- 
\frac{1}{2} \cI_{1/2} (m_{\pi},\bm{B})
\right] 
\notag \\
&& - 
\frac{3}{2  f^2}
\Bigg\{
g_A^2 
\Big[ 
\ol \cJ(m_\pi, \bm{0},0) + \ol \cJ(m_\pi, \bm{B}, 0)
\Big]
- 
\frac{4}{9}  g_{\D N}^2 
\Big[
\ol \cJ ( m_\pi, \bm{0}, \D) 
+ 
\ol \cJ (m_\pi, \bm{B}, \D) 
\Big]
\Bigg\}  
\notag \\
&& +
\frac{3 }{f^2}
\int_0^1 dx 
\Bigg[
g_A^2 \, 
\cJ(m_\pi P_\pi, \bm{0}, \bm{Q},  x \bm{Q}, 0)
-
\frac{4}{9} g_{\D N}^2 \, 
\cJ(m_\pi P_\pi,  \bm{0}, \bm{Q}, x \bm{Q}, \D)
\Bigg],
\end{eqnarray}
}%
with $\bm{Q} = \bm{q} + \bm{B}$. 
The spatial current reads
{\footnotesize
\begin{eqnarray}
\langle \pm | \d J^{+}_{i} | \mp \rangle
&=& 
\frac{1}{f^2} 
\langle \pm | \, [S_{k}, S_j] \, | \mp \rangle
\Bigg\{
2 \delta_{ki}
\left[ 
g_A^2 \, 
\cK^j(m_\pi, \bm{B}, 0)
+ 
\frac{2}{9}
 g_{\D N}^2 \,
\cK^j(m_\pi, \bm{B}, \D)
\right]
\notag \\
&& + 
3 \bm{Q}_k
\int_0^1 dx 
\left[
g_A^2 \, 
\cL^{ji} (m_\pi P_\pi, \bm{0}, \bm{Q},  x \bm{Q}, 0) 
+
\frac{2}{9} g_{\D N}^2 \,
\cL^{ji} (m_\pi P_\pi, \bm{0}, \bm{Q}, x \bm{Q}, \D)
\right]
\Bigg\}
.\end{eqnarray}
}%
We have chosen spin flip matrix elements; 
these are simply related to differences of spin 
polarized matrix elements. 
Because the the finite volume modifications proportional to 
$\cK^j$
are non-vanishing only in the 
directions with non-vanishing twist, 
the spin and momentum transfer structure 
of these terms are identical to the magnetic part of the
current matrix element. 
Consequently one cannot be sensitive to the magnetic
form factor without additionally acquiring finite
volume modifications from $\cK^j$ terms. 
These terms are seen to be numerically larger 
than $\cL^{ji}$, especially for small twists~\cite{Tiburzi:2006px}.
Finally,  
notice these results are independent of the unphysical parameter $g_1$.

\section{Finite Volume Sums} \label{s:FVS}

In this Appendix we describe the evaluation of the mode sums required for the 
finite volume corrections to the nucleon mass and isovector form factors above. 
In the main text, we have used various functions entering in the computation of 
loop graphs in finite volume. Here we evaluate each function systematically 
in terms of Jacobi elliptic functions and error functions.

The basic sums required are of the form
\begin{eqnarray}
\cI_{\b}(\bm{q}, \cM) 
&=& 
\frac{1}{L^3}
\sum_{\bm{n}}\frac{1}{[(\bm{k} + \bm{q})^2 + \cM^2]^\b}
- 
\int \frac{d \bm{k}}{(2 \pi)^3} \frac{1}{[\bm{k}^2 + \cM^2]^\b},
\notag \\
\cI^{i}_{\b}(\bm{q}, \cM) 
&=& 
\frac{1}{L^3}
\sum_{\bm{n}}\frac{\bm{k}^i}{[(\bm{k} + \bm{q})^2 + \cM^2]^\b}
- 
\int \frac{d \bm{k}}{(2 \pi)^3} \frac{\bm{k}^i}{[(\bm{k} + \bm{q})^2 + \cM^2]^\b},
\notag \\
\cI^{ij}_{\b}(\bm{q}, \cM) 
&=& 
\frac{1}{L^3}
\sum_{\bm{n}}\frac{\bm{k}^i \bm{k}^j}{[(\bm{k} + \bm{q})^2 + \cM^2]^\b}
- 
\int \frac{d \bm{k}}{(2 \pi)^3} \frac{\bm{k}^i \bm{k}^j}{[(\bm{k} + \bm{q})^2 + \cM^2]^\b}
\label{eq:Isums}
.\end{eqnarray}
The latter two functions can be derived from the first via differentiation, explicitly the relations are
\begin{equation}
\cI_\b^i(\bm{q},\cM) = 
- 
\frac{1}{2 (\b - 1)} \frac{d}{d \bm{q}^i} \cI_{\b - 1}(\bm{q},\cM)
- 
\bm{q}^i  \cI_\b(\bm{q},\cM)
,\end{equation} 
and
\begin{eqnarray}
\cI_\b^{ij}(\bm{q},\cM) &=& 
\frac{1}{4 (\b - 1)(\b - 2)} \frac{d^2}{d\bm{q}^i d\bm{q}^j} \cI_{\b-2}(\bm{q}, \cM)
\notag \\
&& + 
\frac{1}{2 (\b - 1)}
\left( \d^{ij} + \bm{q}^i \frac{d}{d \bm{q}^j} + \bm{q}^j \frac{d}{d\bm{q}^i} \right) 
\cI_{\b-1}(\bm{q}, \cM)
+
\bm{q}^i \bm{q}^j \cI_\b(\bm{q}, \cM)
.\end{eqnarray}
Evaluating the first function in Eq.~\eqref{eq:Isums} for arbitrary $\b$, we find
\begin{equation}
\cI_\b(\bm{q}, \cM) = \frac{1}{8 \pi^{3/2} \Gamma(\b)} \int_0^\infty d \tau \, \tau^{\b - \frac{5}{2}} e^{-\tau \cM^2} 
\left[
\prod_{j = 1}^{3} \vartheta_3 \left( \bm{q}_j L / 2 , e^{-\frac{L^2}{ 4 \tau}} \right) - 1
\right]
,\end{equation}
where 
$\vartheta_3(q,z)$ 
is a Jacobi elliptic function.

In the main text, we utilized several different mode sums
in the evaluation of finite volume effects.
We now write them out in terms of the basic finite volume functions 
$\cI_\b(\bm{q},\cM)$, $\cI^{i}_\b(\bm{q},\cM)$, and $\cI^{ij}_\b(\bm{q},\cM)$. 
With the abbreviation $\b_\D^2 = m^2 + 2 \l \D + \l^2$, specifically we have
\begin{eqnarray}
\cJ(m, \bm{A},\bm{B}, \bm{C}, \D) 
&=&  
\int_0^\infty d \l \, \l
\left[
\delta^{ij} \cI^{ij}_{5/2} (\bm{C}, \b_\D)
+ 
(\bm{A} + \bm{B})^i 
\cI^i_{5/2} (\bm{C}, \b_\D)
+ 
\bm{A} \cdot \bm{B} \, 
\cI_{5/2} (\bm{C}, \b_\D)
\right],
\notag \\
\\
\cJ^j (m, \bm{A}, \bm{B}, \D) 
&=& 
\int_0^\infty d\l \, \l 
\left[
\cI^j_{5/2} (\bm{B}, \b_\D)
+ 
\bm{A}^j 
\cI_{5/2} (\bm{B}, \b_\D)
\right] ,
\\
\cK(m, \bm{A},\D) 
&=&  
\int_0^\infty d \l
\left[
\delta^{ij} \cI^{ij}_{3/2} (\bm{A}, \b_\D)
+ 
2 \bm{A}^i \cI^i_{3/2} (\bm{A}, \b_\D)
+ 
\bm{A}^2
\cI_{3/2} (\bm{A}, \b_\D)
\right],
\\
\cK^j (m, \bm{A}, \D) 
&=& 
\int_0^\infty d\l
\left[
\cI^j_{3/2} (\bm{A}, \b_\D)
+ 
\bm{A}^j 
\cI_{3/2} (\bm{A}, \b_\D)
\right], \\
\cL^{ij}(m, \bm{A}, \bm{B}, \bm{C}, \D) 
&=& 
\int_0^\infty d\l 
\Big[
2 \cI^{ij}_{5/2} (\bm{C}, \b_\D)
+
2 \bm{A}^i
\cI^{j}_{5/2}  (\bm{C}, \b_\D)
\notag \\
&& \phantom{space}
+
(\bm{A} + \bm{B})^j 
\cI^{i}_{5/2} (\bm{C}, \b_\D)
+ 
\bm{A}^i ( \bm{A} + \bm{B})^j \,
\cI_{5/2}  (\bm{C}, \b_\D)
\Big] 
.\end{eqnarray}

Lastly the evaluation of the $\l$-integrals can be done in closed form. 
For completeness the required $\l$-parameter integrals are
\begin{eqnarray}
\int_0^\infty d \l \, e^{- \tau (\l^2 + 2 b \l + c^2)} &=& \frac{1}{2} \sqrt{\frac{\pi}{\tau}} e^{\tau (b^2 - c^2)} \Erfc (b \sqrt{\tau}),
\\
\int_0^\infty d \l \, \l \,  e^{- \tau (\l^2 + 2 b \l + c^2)} &=& \frac{1}{2} \sqrt{\frac{\pi}{\tau}} e^{-\tau c^2 }
\left[ 
\frac{1}{\sqrt{\pi \tau}} - b e^{\tau b^2} \Erfc (b \sqrt{\tau})
\right]
,\end{eqnarray}
where $\Erfc (x) = 1 - \Erf (x)$, and $\Erf (x)$ is the standard error function.

\bibliography{hb}

\end{document}